\documentclass{article}
\usepackage{graphicx}
\usepackage[round]{natbib}
\usepackage{epsfig}
\usepackage{longtable}
\usepackage{subfigure}
\title{
Predicting basin and landfalling hurricane numbers from sea surface temperature
 }

\author{
Stephen Jewson (RMS)\footnote{\emph{Correspondence email}: \texttt{stephen.jewson@rms.com}}\\
Roman Binter (LSE)\\
Shree Khare (RMS)\\
Kechi Nzerem (RMS)\\
Adam O'Shay (RMS)\\
}

\begin{document}
\maketitle

\begin{abstract}
We are building a hurricane number prediction scheme based on first predicting main development region
sea surface temperature (SST),
then predicting the number of hurricanes in the Atlantic basin given the SST prediction, and finally predicting
the number of US landfalling hurricanes based on the prediction of the number of basin
hurricanes.
We have described a number of SST prediction methods in previous work.
We now investigate the empirical relationship between SST and basin hurricane numbers, and
put this together with the SST predictions
to make predictions
of both basin and landfalling hurricane numbers.
\end{abstract}

\section{Introduction}

We are interested in developing practical methods for the
prediction of the distribution of the number of hurricanes that might make landfall
in the US over the next 5 years.
We are investigating a number of ways that one might make such a prediction, such as methods
based on a change point analysis of the historical hurricane record
(see~\citet{e01}), and methods based on
predictions of sea surface temperatures (SST).
This article contributes to the development of one particular 3 step SST method that works as follows:

\begin{enumerate}

    \item We predict the distribution of possible SSTs in the main development
    region (MDR) over the next 5 years

    \item We predict the distribution of the number of hurricanes in the Atlantic, given the
    SST forecast from step 1, using a model for the relationship between MDR SST and Atlantic hurricane numbers

    \item We predict the distribution of the number of hurricanes making US landfall, given the prediction
    for the number in the Atlantic from step 2, and a model for the relationship between Atlantic hurricane
    numbers and numbers of hurricanes making landfall in the US

\end{enumerate}

The first step, how to predict MDR SST, has been considered in~\citet{j92}
and~\citet{e20}. We now consider steps 2 and 3, and make some predictions.
That there is a
non-trivial relationship between SST and the number of Atlantic
basin hurricanes is both physically intuitive and is supported by a
number of studies in the meteorological literature.
For instance, \citet{peixoto} discuss how long-term variability in hurricane
activity is ultimately driven by the ocean, with its large thermal
and mechanical inertia. Also, ocean SSTs play a direct role in providing
energy to developing tropical cyclones \citep{landsea98,
saunders97}, and higher SSTs decrease the stability of the atmosphere,
making tropical cyclones more resistant to windshear
\citep{demaria96}. Much of this has been known for a long time: \citet{gray68} discusses how warm SSTs promote
tropical cyclone development and enhancement. The statistical
relationship between Atlantic hurricane activity and SSTs has been
studied by a number of authors (e.g. \citet{shapiro82, shapiro89,
saunders97, goldenberg, landsea99}) and is a key aspect in a number
of recent studies concerning the impacts of long-term climate
trends on hurricane frequency and intensity (e.g.
\citet{emanuel05, kerr05, trenberth05, webster05, klotzbach06,
sriver06, elsner06}).

As this is our first attempt to relate SST to
hurricane numbers, and our first attempt to build a forecasting
system on that basis, we are keen to keep our models as simple as
possible. For this reason, we restrict ourselves to representing
SST variability using just a single index, and we choose what we
think is the simplest possible reasonable index: the average of
summer SST in the MDR region of the tropical Atlantic.
Based on an analysis of correlation between hurricane numbers and MDR SSTs
in different months we define summer to be July to September.
In the future it may be appropriate to extend our prediction
system to consider more than one SST index, and to derive our
index or indices on the basis of some kind of regression or
eigenvector analysis.

Given our SST index, our first goal is to test a number of statistical models that relate this index
to the observed number of hurricanes. We consider both the total number of hurricanes, and the number
of intense hurricanes, where we define intense as falling in categories 3-5. We fit our statistical
models to data from both 1900-2005 and 1950-2005, for reasons discussed below.

Having settled on a couple of models for the SST to hurricane number relationship,
we then convert SST predictions derived from~\citet{e20} into hurricane number
predictions for 2006-2010.

The rest of this article proceeds as follows.
In section~\ref{data} we briefly discuss the data we use for this study.
In section~\ref{models} we present the statistical models for the SST-hurricane number relationship that we will test.
In section~\ref{results} we show the results from these tests,
and in section~\ref{summary} we summarise and discuss those results.
Then in section~\ref{p1} we describe how we can use these models to make hurricane number predictions,
and in sections~\ref{p2} and~\ref{p3} we make some predictions,
for all hurricanes and intense hurricanes, respectively.
Finally in section~\ref{conc} we conclude.

\section{Data}\label{data}

We use two data sets for this study: one for SST, and the other for hurricane numbers.

The SST data we use is HadISST, available
from~\texttt{http://www.hadobs.org}, and described in
\citet{hadisst}. From this data we create an annual index
consisting of the average SST in the region (10$^o$-20$^o$N, 15$^o$-70$^o$W) for
the months July to September. This is exactly the same index as
was used in our SST prediction study described in~\citet{e20}.

The hurricane number data is derived from the standard HURDAT data
base, available \newline
from~\texttt{http://www.aoml.noaa.gov/hrd/hurdat}, and described
in \citet{hurdat}.

In both cases we consider the data to be exact, i.e. without observational errors. This is
pragmatic, since neither data-set is delivered with any estimates of the likely error. However,
given the (widely discussed) possibility that both data sets may be less accurate earlier in the
time period we repeat our analyses for both the periods 1900-2005 and 1950-2005.

\section{Statistical models for the SST to hurricane number relationship}\label{models}

The purpose of sections~\ref{models}, \ref{results} and~\ref{summary} of this article is to compare the performance of a number of simple
statistical models for modelling the relationship between our MDR SST index and observed
hurricane numbers. The models we consider are as follows:

\subsection{Models}

\subsubsection{Poisson distribution independent of SST}

The simplest model we consider models the number of hurricanes as a poisson distribution independent of SST.
We include this model as a baseline that the subsequent models should be able to beat.

We write this model as:
\begin{equation}
 n \sim \mbox{Po}(\mbox{rate}=\alpha)
\end{equation}

where $n$ is the number of hurricanes.
We estimate the parameter $\alpha$ as the mean of the observed hurricane numbers.

\subsubsection{Linear normal regression}

Our next model is plain-vanilla linear regression, with both the residuals and the response
modelled as normal distributions.

We write this model as:
\begin{equation}
 n \sim \mbox{N}(\alpha+\beta s, \sigma^2)
\end{equation}

where $s$ is the SST index. We fit the parameters using maximum likelihood, which is equivalent
to least squares in this case.

From the point of view of making point predictions of future hurricane numbers this is a
perfectly reasonable first model, and the prediction is $\alpha+\beta s$. From the point of view
of making a probabilistic prediction of future hurricane numbers this model is slightly odd,
since it models a non-negative integer (the number of hurricanes) using a real number that
can be both positive and negative. Nevertheless we include this model for two reasons:
(a) as another baseline for comparison, since it is simple and well understood, and there is
a closed-form solution for the parameters, and (b)
since this model can easily be optimised for predictive purposes (the next model).

\subsubsection{Damped linear normal}

Our third model is an adaption of linear regression. Standard linear regression, when fitted using
in-sample fitting techniques such as maximum likelihood or least squares, is an inherently overfitted
model. In other words, these fitting techniques optimise the model's ability to predict the training
data, but not to extrapolate. For large samples and strong signals this doesn't matter, but for
smaller samples and weak signals this means that linear regression does not give good predictions, even
when the process being predicted really does consist of a linear trend plus gaussian noise. Adapting linear regression
so that it is optimised to make good predictions is, however, difficult, and is not guaranteed to be
successful. Our third model is one such simple adaption, due to~\citet{j58}, in which the slope of the trend
is reduced in order to reduce the overfitting. This model may or may not make better predictions than
the unadapted linear regression model
because of the difficulty of estimating exactly what this slope reduction should be.

We write this model as:
\begin{equation}
 n \sim \mbox{N}(\alpha+k \beta s, \sigma^2)
\end{equation}

We fit this model by first fitting the underlying linear regression model, and then calculating the
adjustment $k$ using the expressions in~\citet{j58}.

As with the linear normal model this model is a reasonable model if used to make point predictions,
but is a slightly odd model to use for probabilistic predictions, for the same reasons as given above.
We include it
in order to see the extent to which overfitting may be an issue.

\subsubsection{Linear Poisson}

The two normally distributed models given above can be criticized, as probabilistic forecast models,
for using a distribution that is clearly not close to being correct. To overcome this criticism
we now change the distribution from normal to poisson. We write this model as:
\begin{equation}
n \sim \mbox{Po} (\mbox{rate}=\alpha+\beta s, \sigma^2)
\end{equation}

We fit this model using iteratively reweighted least squares.

At a mathematical level this model might be criticized because for certain values of the parameters
the poisson rate can be negative. However we have found that this is not a problem for the data and parameter ranges
that are of interest to us.

The standard fitting procedure we use for this model is an in-sample fitting procedure, and
thus leads to inherently overfitted parameter estimates, and sub-optimal predictive properties.
It would be possible in principle to fit a `damped' version of this model, in the same way that we have fitted
a damped version of linear regression, in order to attempt to overcome this problem.
However there appears to be no simple analytic way to do this,
and one would have to use more complex methods such as cross-validation. That is beyond the scope
of this study, but might be an interesting avenue for future work.

\subsubsection{Exponential poisson}

The next model that we use is the same as the previous model, but exchanges a linear relationship
for the rate of the poisson distribution for an exponential one. This model is common in the statistical
literature, where it is known as either `poisson regression' or
`a generalised linear model for the poisson distribution with a log-linear link function'.

We write this model as:
\begin{equation}
n \sim \mbox{Po} (\mbox{rate}=\mbox{exp}(\alpha+\beta s))
\end{equation}

At a mathematical level this model avoids the potential problem with negative rates described above, since the rates are positive for all parameter
values and data sets. This model has previously been used to model the relationship between SST and
hurricane numbers by~\citet{elsners93} and~\citet{solow00}.

Comparing the linear poisson and exponential poisson models, one obvious question is, is there any statistical evidence
for the non-linearity included in the exponential poisson model? We will discuss this question when we present
our results below.

\subsubsection{Exponential negative binomial}

Our final model is an adaption of the previous model that allows the distribution around the
mean to be negative binomial rather than exponential, and thus has one extra parameter that allows
for the variance to be different from the mean.

We write this model as:
\begin{equation}
n \sim \mbox{NB} (\mbox{mean}=\alpha+\beta s,\mbox{variance}=\gamma)
\end{equation}

\subsection{Model comparison}

How are we going to compare the results from these different
models? We consider two scores, one of which assesses the ability
of the models to make point predictions, and the other of which
assesses the ability of the models to make probabilistic
predictions. For each of these scores we consider in-sample
scores, which are not really what we care about, but are included
for interest, and out-of-sample scores, which are the real test.
The out-of-sample scores are calculated using leave-one-out
cross-validation, a.k.a. the Quenouille-Tukey jack-knife
\citep{quenouille49, tukey58}.

We score the point predictions using the most obvious score available: the root mean
square error. We score the probabilistic predictions using (what we consider to be) the
most obvious probabilistic score, which is the mean out-of-sample log-likelihood.

\section{Comparing the performance of the statistical models}\label{results}

We now present results from our comparisons of the various models.
First we consider models for the total number of hurricanes, for the periods 1900-2005 and
1950-2005, and then we consider models for intense hurricane numbers for the same two periods.
Correlations between SST and hurricane numbers for these 4 cases are shown in table 1.

In each of the 4 cases we produce a standard set of diagnostics, consisting
of four tables and three graphs. The four tables are:

\begin{itemize}

    \item the scores achieved by the models

    \item the fitted parameter values, including standard error estimates based on the Fisher information

    \item the percentage of times each model wins in pairwise comparisons, and corresponding p-values for point predictions

    \item pairwise comparisons and corresponding p-values for probabilistic predictions

\end{itemize}

The graphs are:
\begin{itemize}
    \item a scatter plot showing the data on which the model is based
    \item the same scatter plot, showing the decade in which each data point occurred
    \item the same scatter plot showing the fitted curves from the six models
\end{itemize}


\subsection{All hurricanes, 1900-2005}

The first results we present are based on all hurricanes, and data
from 1900 to 2005. The scatter plot shown in figure~1 shows a
clear relationship between the SSTs and the hurricane numbers
during this period, with warmer SSTs coinciding with more
hurricanes. By eye, the relationship looks more or less linear.
The linear correlation was found to be 0.56, while the rank
correlation was found to be 0.51, as shown in table~1.

The decade scatter plot is shown in figure~2. Note that the most
recent period labelled by `A' gives relatively high SST and
basin number count.

Table~2 shows the score comparisons for the six models for this
data set. Considering the RMSE scores we see that all the
non-trivial models comfortably beat the trivial flat-line model,
as we'd guess would be the case from the scatter plot. Table~4,
which shows results for pairwise comparisons in RMSE, shows that
the differences between the trivial model and the non-trivial
models are statistically significant (at the 5 percent level),
with the exception of the exponential binomial model.

The differences between the performance of the 5 non-trivial
models are small. The exponential models yield the best
out-of-sample RMSEs, but none of the RMSE differences between the
five models are statistically significant. Given that the
exponential poisson model gives the second lowest out-of-sample
RMSE and beats the flat poisson model in a statistically
significant way, one might choose the exponential poisson model as
the best one. However, since the non-trivial models are not
different in a statistically significant way, one shouldn't be
surprised if that result were overturned given more data.


These models all explain around 30\% of the variance in the
hurricane number time series.

Considering the out of sample log-likelihood scores in table~2, we
find that the linear and exponential poisson are better than the
flat poisson model in a statistically significant way. The linear
and damped linear normal are worse than the flat poisson model in
a statistically significant way. The exponential negative binomial
does yield the best out-of-sample log-likelihood score, but the
result is not statistically significant. Based on the
log-likelihood scores, one might choose the exponential poisson
model once again.

The parameters of all the models are reasonably well estimated:
all the parameters given in table~3 are significantly different
from zero (judging by the standard error estimates, and assuming
normality for the sampling distributions). For instance, the
slopes in the linear models are between 4.3 and 4.6, with standard
error of roughly 0.7. Each extra degree of SST is therefore
related to just over 4 more hurricanes, plus or minus 1.5
hurricanes. The in-sample fits to the data for the various models
are displayed in figure~3.

The damping parameter in the damped linear trend model is very close to one.
This suggests that the models are not significantly overfitted, and there is no real need to
use such damped models in this case.
This is because the signal is strong enough that we can estimate it reasonably well.

\subsection{All hurricanes, 1950-2005}

Given the doubts that one might have about the quality of both the
SST and the hurricane number data prior to 1950, it makes sense to
repeat the analysis given in the previous section for just the
more recent data from 1950 to 2005. The corresponding results and
data plots are provided in tables~6, 7, 8 and 9 and figures~4, 5
and 6. For this data set, a linear correlation of 0.62 and rank
correlation of 0.56 was found.

There are only small differences in the results relative to the analysis
based on 1900-2005 data. Once again, the
non-trivial models all beat the trivial constant level model.
These results are statistically significant for the linear normal,
damped linear normal and linear poisson models. The differences
between the non-trivial models are not, once again, statistically
significant. Based on these results, one may be inclined to choose
the linear poisson model, as it yields the lowest RMSE of the
models that beat the flat poisson model in a statistically
significant way.

These models all explain around 40\% of the variance in the
hurricane number time series (a little higher than before).

With regards to the log-likelihood scores, the linear and
exponential poisson models and the exponential negative binomial
models beat the flat poisson model, whereas the linear normal and
damped linear normal do not. The result for the linear poisson
model is statistically significant. The linear and damped linear
normal are defeated by the flat poisson model in a statistically
significant way. Given these results, one might choose the linear
poisson model once again. However, it is possible
that this conclusion would be overturned by using more data as the
differences between the linear poisson, exponential poisson and
exponential negative binomial are not statistically significant.

The slope parameters are again significantly different from zero,
but the slope of the linear relations is now a bit higher, giving between
5.0 and 5.4 hurricanes per degree, with a slightly larger
uncertainty of around 1.0 hurricanes per degree. Given the
uncertainties, the slope estimates from the two data sets are
entirely consistent.

\subsection{Intense hurricanes, 1900-2005}

We now consider the relationship between MDR SST and the number of
intense hurricanes. Considering the scatter plot in figure~7, and
comparing with the scatter plot in figure~1, we see immediately
that the relationship is less clear than before. The linear
correlation for this data set is 0.52 and the rank correlation is
0.54. This may not be because the underlying relationship is any
less strong: simple statistical arguments suggest that the
relationship will appear less strongly in the data just because
there are fewer events.

We now consider the results in tables~10, 11, 12 and 13. The
non-trivial models all beat the trivial constant level model
statistically significantly, but are not statistically
significantly different from each other in terms of the RMSE
scores. The parameters of all models are significantly different
from zero, and the linear models give roughly 2.9 extra
hurricanes per extra degree of SST, with a standard error of
around 0.4. The models explain around 25\% of the variability in
the number of intense hurricanes.

As far as the log-likelihood scores are concerned, the linear
normal and damped linear normal are defeated by the flat poisson
model in a statistically significant way. The linear and
exponential poisson models and the negative binomial model defeat
the flat poisson model in a statistically significant way. The
differences between the linear poisson, exponential poisson and
exponential negative binomial are not statistically significant.

\subsection{Intense hurricanes, 1950-2005}

Finally we consider intense hurricane numbers for the more recent
data, for which results are shown in tables 14, 15, 16 and 17 and figures~10 through~12.
Once again, the non-trivial models defeat the flat poisson
model in a statistically significant way. The differences among
the non-trivial models in this case are not statistically
significant. Once again, the parameter estimates for the models
appear to be significantly different from zero. The linear models
give a slightly higher number of hurricanes per degree, roughly
around 3.4 extra hurricanes per extra degree, with standard error
of around 0.7.

With regards to the log-likelihood scores, the linear normal and
damped linear normal are defeated by the flat poisson model in a
statistically significant way. The linear poisson and exponential
negative binomial defeat the flat poisson model in a statistically
significant way, whereas the exponential poisson does not. The
differences between the linear poisson, exponential poisson and
exponential negative binomial are not statistically significant.

\section{Summary of statistical model results}\label{summary}

In sections~\ref{models} and~\ref{results}
we have considered how to model the relationship between MDR SST and the number of hurricanes in the
Atlantic basin. We considered both the total number of hurricanes and the number of intense hurricanes.
We now summarise the results of this investigation.

W.r.t. the total number of hurricanes our findings are:
\begin{itemize}

    \item that there is a clear and statistically significant relationship, such that higher SSTs
    correspond to higher numbers of hurricanes, with one degree of SST relating to between 4.0 and 5.5 extra
    hurricanes

    \item using only more recent data the relationship is slightly stronger, but is less accurately estimated

    \item that statistical models of this relationship give better point
    predictions of hurricane numbers than a simple model that ignores this relationship

    \item w.r.t. point predictions, all the non-trivial models
    defeat the flat poisson model in a statistically significant
    way with the exception of the exponential poisson and
    exponential negative binomial
    model. The differences between the non-trivial models are,
    however, not statistically significant. If one \emph{had} to
    choose among the models based on the RMSE results, one might
    choose the linear poisson model as it yields the lowest
    RMSE of the models that beat the trivial model in
    statistically significant way.


    \item that the non-trivial models explain between 29\% and 44\%
     of the variability in the number of hurricanes

    \item w.r.t probabilistic predictions, the linear and damped
    linear normal models are defeated by the flat poisson model in
    a statistically significant way. The linear poisson model
    defeats the flat-line model in a statistically significant
    way.

\end{itemize}

That there is a non-trivial relationship between Atlantic MDR SSTs
and basin-wide hurricane activity is in general agreement with
statistical analyses performed in both \citet{klotzbach06} and
\citet{landsea99}.

W.r.t the number of intense hurricanes our findings are the same,
with the exception of:
\begin{itemize}
    \item each extra degree of SST gives between 3.0 and 3.5 extra intense hurricanes
    \item the non-trivial models explain slightly less of the variability in the numbers of hurricanes: only 25\%
    \item w.r.t. the point predictions, all the non-trivial models
    defeat the flat-line model in a statistically significant
    way.

    \item w.r.t. the probabilistic predictions, the linear poisson
    and negative binomial models defeat the flat poisson model
    in a statistically significant way for both data sets.

    \end{itemize}

Our results for intense hurricanes are broadly consistent with an
analysis done by \citet{hoyos06} which suggests that increasing
number of category 4-5 hurricanes is directly linked to the trend
in tropical SSTs.

Given all of this, what models would we recommend to use to model the SST
to hurricane relationship? Firstly, it probably makes sense to use only
the recent data, since it is possible to estimate the regression parameters
reasonbly well using this data, and it avoids doubts about data quality.
Assuming we need probabilistic forecasts of hurricane activity, then we
have seen that the normal distribution models don't work well.
This leaves 3 reasonable models: linear poisson, exponential poisson,
and negative binomial. We could elimatine the negative binomial
model on the basis that it is more complex than the other two, but performs
no better. This then leaves us with the linear poisson and exponential poisson models.
On the basis of our results it is not possible to distinguish between these models,
and this is perhaps the most important result of this paper: even though previous
authors have, by default, used an exponential poisson model for the SST to hurricane
number relationship, the linear poisson model works just as well, and in some
forecast applications (especially those that involve applying the modelled
SST-hurricane relationships for extreme values of the SST) is likely to give quite different results.

We end this section by mentioning that a number of studies suggest
that the strength of the relationship between SST and hurricane
frequency is dependent on the region of the north Atlantic being
considered \citep{shapiro89, raper92, goldenberg}. Understanding the regional
dependence of the statistical relationship between SST and hurricane
numbers is therefore an
interesting avenue for future work.

\section{Making SST based predictions of hurricane numbers}\label{p1}

We now have all the pieces we need to make SST-based predictions
of hurricane numbers.
Firstly,
in~\citet{e20}, we have derived simple statistical methods for
predicting SST. We will take three SST
predictions from that article: the flat-line model based on 8
years of data (FL), the linear trend model based on 24 years of
data, and a damped linear trend model that is the mean of these
two.

Secondly, in sections~\ref{models} and~\ref{results} above,
we have analysed the relationship
between SST and the number of hurricanes in the Atlantic basin. We
were able to find two relatively simple models that were
significantly better than the trivial model of no relationship in
out-of-sample tests. The first of these models represents the mean
number of basin hurricanes as a linear function of SST, and the
distribution as a poisson distribution. The second model
represents the mean number of basin hurricanes as an exponential
function of SST, and the distribution once again as poisson. A
more complex model with more parameters (that represents the
distribution as negative binomial) was not significantly better,
so we ignore it.

Thirdly, using historical hurricane data for the period 1950 to
2005 we can estimate the probability that individual hurricanes
make landfall in the US. For cat 1-5 hurricanes the estimate of
this probability is 0.254 while for cat 3-5 hurricanes the
estimate is 0.240.


Finally, in~\citet{e06}, we have derived simple analytic
relationships that allow us to put this all together and predict
the mean, variance, and standard error of the number of
landfallling hurricanes as a function of the mean, variance and
standard error of an SST forecast. In particular, we use the
equations given in section 9 of that paper.

Combining our three SST models with two ways of converting SST to
basin hurricane numbers gives a total of six different forecast
methods. We present results for all of these six different
methods, since the SST forecasts capture a range of possible
points of view about the possible future behaviour of SST, and the
two SST-to-basin relationships are both sensible models, and can't
be distinguished given the observational data, but give clearly
different final answers.

\section{Predictions for category 1-5 landfalling hurricanes}\label{p2}

We now present our various predictions for SST and
the number of category 1-5 hurricanes
at US landfall.

First, in figure~\ref{f01}, we show the three SST predictions.

Second, in tables~\ref{b01} to \ref{b05}, we show details of the
six predictions of \emph{basin} hurricane numbers that we derive
from these SST predictions. Since these are just an intermediate
step in the process of predicting landfalling hurricane numbers,
we don't discuss these in detail. We note
briefly that the predictions for individual years range from 8.0
to 10.5 hurricanes per year, and the 5 year averages range from
8.9 to 9.9 hurricanes per year.

Thirdly, in tables~\ref{b06} to \ref{b10}, we show details of the
six predictions of \emph{landfalling} hurricane numbers that we
derive from these predictions of basin hurricane numbers by
multiplying by the estimated probabilities that a hurricane will
make landfall. These predictions are also illustrated in
figures~\ref{f02} to \ref{f07}. Model 1 gives a flat prediction of
future hurricane numbers by year since it is based on a flat
prediction of SST and a linear conversion model. Model 4 gives a
very gradual increase in the prediction of the mean number of
hurricanes with lead time because of the increasing uncertainty of
the SST prediction with lead time, in combination with the
non-linearity of the conversion model. Models 3 and 6 give rapidly
increasing predictions of hurricane numbers since they are based
on rapidly increasing SST predictions. Models 2 and 5 lie
somewhere in between these two extremes. Models 4, 5 and 6, based
on the exponential poisson model for the relation between SST and
hurricane numbers, all give higher predictions of future hurricane
numbers than models 1, 2 and 3 that are based on the linear
poisson relation.

%
%

\section{Predictions for category 3-5 landfalling hurricanes}\label{p3}

For reference we also include predictions for category 3-5
landfalling hurricanes, in tables~\ref{b11} to~\ref{b20}.

\section{Conclusions}\label{conc}

We have described a hurricane number prediction scheme based on three
steps. In step 1 we predict MDR
SSTs, in step 2 we predict basin hurricane numbers given the MDR
SST predictions, and in step 3 we predict landfalling hurricane
numbers given the prediction of basin hurricane numbers.
We have tested a number of different models for the relationship
between SST and hurricane numbers. We have found 2 models that beat the
other models tested, but we can't say which of the two is better based
only on the data.
We make 6 predictions of landfalling hurricane numbers by combining 3
different SST predictions with these 2 different methods for converting
SSTs to basin hurricane numbers. The SST prediction models,
which are taken from~\citet{e20}, range
from a model in which SST remains at a constant level to a model
in which the SST increases rather rapidly.
Finally we convert the basin number predictions to predictions of numbers
of landfalling hurricanes using a constant probability of landfall
estimated from 56 years of historical data.
Putting this all together,
the averages of the
predictions for the number of landfalling hurricanes
over the next 5 years range from 2.04 to 2.52
hurricanes per year.
These predictions are broadly similar to predictions of
hurricane numbers from time-series based methods such as those we
have described in~\citet{e01}, but extend to higher numbers of
hurricanes for the highest predictions.

Finally we note that there is a large spread between the
predictions based on different SST forecasts. Reducing the
uncertainty as to how to predict future SSTs would perhaps be the
easiest way for us to reduce the uncertainty around our future
hurricane number predictions.


\bibliography{../../bib/jewson,../../bib/shreeonly,../../bib/shreetwo}

\begin{thebibliography}{29}
\providecommand{\natexlab}[1]{#1}
\providecommand{\url}[1]{\texttt{#1}}
\expandafter\ifx\csname urlstyle\endcsname\relax
  \providecommand{\doi}[1]{doi: #1}\else
  \providecommand{\doi}{doi: \begingroup \urlstyle{rm}\Url}\fi

\bibitem[Binter et~al.(2006)Binter, Jewson, Khare, O'Shay, and Penzer]{e01}
R~Binter, S~Jewson, S~Khare, A~O'Shay, and J~Penzer.
\newblock {Year ahead prediction of US landfalling hurricane numbers: the
  optimal combination of multiple levels of activity since 1900}.
\newblock \emph{arXiv:physics/0611070}, 2006.
\newblock RMS Internal Report E01.

\bibitem[Demaria(1996)]{demaria96}
M~Demaria.
\newblock The effect of vertical shear on tropical cyclone intensity change.
\newblock \emph{Journal of Atmospheric Sciences}, 53:\penalty0 2076--2088,
  1996.

\bibitem[Elsner(2006)]{elsner06}
J~Elsner.
\newblock {Evidence in support of the climate change - Atlantic hurricane
  hypothesis}.
\newblock \emph{Geophysical Research Letters}, 33, 2006.

\bibitem[Elsner and Schmertmann(1993)]{elsners93}
J~Elsner and C~Schmertmann.
\newblock {Improving extended-range seasonal predictions of intense Atlantic
  hurricane activity}.
\newblock \emph{{Weather and Forecasting}}, 3:\penalty0 345--351, 1993.

\bibitem[Emanuel et~al.(2005)Emanuel, S, E, and C]{emanuel05}
K~Emanuel, Ravela S, Vivant E, and Risi C.
\newblock A combined statistical-deterministic approach of hurricane risk
  assessment.
\newblock Unpublished manuscript, 2005.

\bibitem[Goldenberg et~al.(2001)Goldenberg, Landsea, Mestas-Nunez, and
  Gray]{goldenberg}
S~Goldenberg, C~Landsea, A~Mestas-Nunez, and W~Gray.
\newblock {The recent increase in Atlantic hurricane activity: causes and
  implications}.
\newblock \emph{Science}, 293:\penalty0 474--479, 2001.

\bibitem[Gray(1968)]{gray68}
W~M Gray.
\newblock {Global view of the origin of tropical disturbances and storms}.
\newblock \emph{{Monthly Weather Review}}, 95:\penalty0 55--73, 1968.

\bibitem[Hoyos et~al.(2006)Hoyos, Agudelo, Webster, and Curry]{hoyos06}
C~D Hoyos, P~A Agudelo, P~J Webster, and J~A Curry.
\newblock {Deconvolution of the factors contributing to the increase in global
  hurricane intensity}.
\newblock \emph{Science Express}, 2006.

\bibitem[Jarvinen et~al.(1984)Jarvinen, Neumann, and Davis]{hurdat}
B~Jarvinen, C~Neumann, and M~Davis.
\newblock {A tropical cyclone data tape for the North Atlantic Basin,
  1886-1983: Contents, limitations, and uses}.
\newblock Technical report, {NOAA Technical Memorandum NWS NHC 22}, 1984.

\bibitem[Jewson(2007)]{e06}
S~Jewson.
\newblock {Predicting Hurricane Numbers from Sea Surface Temperature: closed
  form expressions for the mean, variance and standard error of the number of
  hurricanes}.
\newblock \emph{arXiv:physics/0701167}, 2007.
\newblock RMS Internal Report E06.

\bibitem[Jewson and Penzer(2004)]{j58}
S~Jewson and J~Penzer.
\newblock Optimal year ahead forecasting of temperature in the presence of a
  linear trend, and the pricing of weather derivatives.
\newblock \emph{http://ssrn.com/abstract=563943}, 2004.

\bibitem[Kerr(2005)]{kerr05}
R~A Kerr.
\newblock {Atlantic climate pacemaker for millenia past, decades hence?}
\newblock \emph{Science}, 309:\penalty0 41--43, 2005.

\bibitem[Klotzbach(2006)]{klotzbach06}
P~J Klotzbach.
\newblock {Trends in global tropical cyclone activity over the past twenty
  years}.
\newblock \emph{Geophysical Research Letters}, 33:\penalty0 1--4, 2006.

\bibitem[Laepple et~al.(2007)Laepple, Jewson, Meagher, O'Shay, and Penzer]{e20}
T~Laepple, S~Jewson, J~Meagher, A~O'Shay, and J~Penzer.
\newblock {Five-year ahead prediction of Sea Surface Temperature in the
  Tropical Atlantic: a comparison of simple statistical methods}.
\newblock \emph{arXiv:physics/0701162}, 2007.

\bibitem[Landsea et~al.(1999{\natexlab{a}})Landsea, Bell, and Gray]{landsea98}
C~W Landsea, G~D Bell, and W~M Gray.
\newblock The extremely active 1995 atlantic hurricane season: Environmental
  conditions and verification of seasonal forecasts.
\newblock \emph{Monthly Weather Review}, 126:\penalty0 1174--1193,
  1999{\natexlab{a}}.

\bibitem[Landsea et~al.(1999{\natexlab{b}})Landsea, Jr, Mestas-Nunez, and
  Knaff]{landsea99}
C~W Landsea, R~A~Pielke Jr, A~M Mestas-Nunez, and J~A Knaff.
\newblock Atlantic basin hurricanes: Indices of climate changes.
\newblock \emph{Climatic Change}, 42:\penalty0 89--129, 1999{\natexlab{b}}.

\bibitem[Meagher and Jewson(2006)]{j92}
J~Meagher and S~Jewson.
\newblock {Year ahead prediction of hurricane season SST in the tropical
  Atlantic}.
\newblock \emph{arXiv:physics/0606185}, 2006.

\bibitem[Peixoto and Oort(1992)]{peixoto}
J~P Peixoto and A~H Oort.
\newblock \emph{Physics of Climate}.
\newblock American Institute of Physics, New York, 1992.

\bibitem[Quenouille(1949)]{quenouille49}
M~Quenouille.
\newblock Approximate tests of correlation in time series.
\newblock \emph{Journal of the Royal Statistical Society, Soc. Series B},
  11:\penalty0 18--84, 1949.

\bibitem[Raper(1993)]{raper92}
S~Raper.
\newblock \emph{Observational data on the relationships between climate change
  and the frequency and magnitude of severe tropical storms: Climate and sea
  level change: Observations, projections and implications}.
\newblock Cambridge University Press, editors: R A Warrick, E M Barrow and T M
  L Wigley, 1993.

\bibitem[Rayner et~al.(2002)Rayner, Parker, Horton, Folland, Alexander, Rowell,
  Kent, and Kaplan]{hadisst}
N~Rayner, D~Parker, E~Horton, C~Folland, L~Alexander, D~Rowell, E~Kent, and
  A~Kaplan.
\newblock {Global analyses of SST, sea ice and night marine air temperature
  since the late nineteenth century}.
\newblock \emph{{Journal of Geophysical Research}}, 108:\penalty0 4407, 2002.

\bibitem[Saunders and Harris(1997)]{saunders97}
M~A Saunders and A~R Harris.
\newblock {Statistical evidence links exceptional 1995 Atlantic hurricane
  season to record sea warming}.
\newblock \emph{Geophysical Research Letters}, 24:\penalty0 1255--1258, 1997.

\bibitem[Shapiro(1982)]{shapiro82}
L~J Shapiro.
\newblock Hurricane climatic fluctuations. part ii: Relation to large-scale
  circulation.
\newblock \emph{Monthly Weather Review}, 110:\penalty0 1007--1013, 1982.

\bibitem[Shapiro and Goldenberg(1989)]{shapiro89}
L~J Shapiro and S~B Goldenberg.
\newblock Atlantic sea surface temperatures and tropical cyclone formation.
\newblock \emph{Journal of Climate}, 11:\penalty0 2598--2614, 1989.

\bibitem[Solow and Moore(2000)]{solow00}
A~R Solow and L~Moore.
\newblock Testing for a trend in a partially incomplete hurricane record.
\newblock \emph{Journal of Climate}, 13:\penalty0 3696--3699, 2000.

\bibitem[Sriver and Huber(2006)]{sriver06}
R~Sriver and M~Huber.
\newblock {Low frequency variability in globally integrated tropical cyclone
  power dissipation}.
\newblock \emph{Geophysical Research Letters}, 33, 2006.

\bibitem[Trenberth(2005)]{trenberth05}
K~Trenberth.
\newblock {Uncertainty in hurricanes and global warming}.
\newblock \emph{Science}, 308:\penalty0 1753--1754, 2005.

\bibitem[Tukey(1958)]{tukey58}
J~W Tukey.
\newblock Bias and confidence in not quite large samples.
\newblock \emph{Annals of Mathematical Statistics}, 29:\penalty0 614, 1958.

\bibitem[Webster et~al.(2005)Webster, Holland, and Chang]{webster05}
P~J Webster, G~J Holland, and H~R Chang.
\newblock {Changes in tropical cyclone number, duration, and intensity in a
  warming environment}.
\newblock \emph{Science}, 309:\penalty0 1844--1846, 2005.

\end{thebibliography}

\newpage

\begin{table}[h!]
\begin{center}
\caption{Linear and Rank Correlations} {\small
\begin{tabular}{|c|c|c|}
\hline
 & Linear Correlation & Rank Correlation \\
\hline
1900 - 2005 Basin vs SST & 0.56 & 0.51 \\
1950 - 2005 Basin vs SST & 0.62 & 0.56 \\
1900 - 2005 Intense Basin vs SST & 0.52 & 0.54 \\
1950 - 2005 Intense Basin vs SST & 0.53 & 0.56 \\
\hline
\end{tabular}
}
\end{center}
\end{table}

\newpage

\begin{table}[h!]
\begin{center}
\caption{RMSE comparison 1900 - 2005 Basin vs SST} {\small
\begin{tabular}{|c|c|c|c|c|c|c|}
\hline
 & model name & RMSE (in) & RMSE (out) & 100-100*RMSE/RMSEconst & LL (in) & LL (out) \\
\hline
model 1 & Flat Poisson & 2.648 & 2.674 & 0 & -2.372 & -2.385 \\
model 2 & Linear Normal & 2.185 & 2.238 & 29.94 & -2.61 & -2.612 \\
model 3 & Damped Linear Normal & 2.185 & 2.239 & 29.867 & -2.61 & -2.612 \\
model 4 & Linear Poisson & 2.187 & 2.233 & 30.24 & -2.169 & -2.189 \\
model 5 & Exponential Poisson & 2.153 & 2.204 & 32.019 & -2.163 & -2.182 \\
model 6 & Exponential Neg. Bin. & 2.153 & 2.168 & 34.249 & -2.163 & -2.171 \\
\hline
\end{tabular}
}
\end{center}
\end{table}
\begin{table}[h!]
\begin{center}
\caption{Model parameters incl. out of sample RMSE 1900 - 2005
Basin vs SST} {\small
\begin{tabular}{|c|c|c|c|c|c|c|c|c|}
\hline
 & $\hat{\alpha}$ & s.e. & $\hat{\beta}$ & s.e. & $k$ & cov & corr & RMSE (out of sample) \\
\hline
model 1 & 1.664 & 0.042 &  &  &  &  &  & 2.674 \\
model 2 & 5.283 & 0.214 & 4.581 & 0.656 &  & 0 & 0 & 2.238 \\
model 3 & 5.283 &  & 4.489 &  & 0.98 &  &  & 2.239 \\
model 4 & 5.283 & 0.223 & 4.34 & 0.662 &  & 0.041 & 0.277 & 2.233 \\
model 5 & 1.625 & 0.044 & 0.849 & 0.127 &  & -0.001 & -0.268 & 2.204 \\
model 6 & 1.625 & 0.044 & 0.849 & 0.127 &  & -0.001 & -0.268 & 2.168 \\
\hline
\end{tabular}
}
\end{center}
\end{table}
\begin{table}[h!]
\begin{center}
\caption{Winning count for particular model 1900 - 2005 Basin vs
SST} {\small
\begin{tabular}{|c|c|c|c|c|c|c|}
\hline
 & model 1 & model 2 & model 3 & model 4 & model 5 & model 6 \\
\hline
model 1 & 0 (1) & 42 (0.968) & 41 (0.98) & 41 (0.98) & 40 (0.987) & 43 (0.928) \\
model 2 & 58 (0.049) & 0 (1) & 45 (0.857) & 46 (0.809) & 52 (0.385) & 45 (0.857) \\
model 3 & 59 (0.032) & 55 (0.191) & 0 (1) & 45 (0.857) & 53 (0.314) & 49 (0.615) \\
model 4 & 59 (0.032) & 54 (0.248) & 55 (0.191) & 0 (1) & 47 (0.752) & 48 (0.686) \\
model 5 & 60 (0.02) & 48 (0.686) & 47 (0.752) & 53 (0.314) & 0 (1) & 43 (0.928) \\
model 6 & 57 (0.103) & 55 (0.191) & 51 (0.461) & 52 (0.385) & 57 (0.103) & 0 (1) \\
\hline
\end{tabular}
}
\end{center}
\end{table}
\begin{table}[h!]
\begin{center}
\caption{Winning count (LL) for particular model 1900 - 2005 Basin
vs SST} {\small
\begin{tabular}{|c|c|c|c|c|c|c|}
\hline
 & model 1 & model 2 & model 3 & model 4 & model 5 & model 6 \\
\hline
model 1 & 0 (1) & 77 (0) & 77 (0) & 41 (0.98) & 41 (0.98) & 44 (0.897) \\
model 2 & 23 (1) & 0 (1) & 45 (0.857) & 12 (1) & 11 (1) & 11 (1) \\
model 3 & 23 (1) & 55 (0.191) & 0 (1) & 12 (1) & 11 (1) & 11 (1) \\
model 4 & 59 (0.032) & 88 (0) & 88 (0) & 0 (1) & 47 (0.752) & 48 (0.686) \\
model 5 & 59 (0.032) & 89 (0) & 89 (0) & 53 (0.314) & 0 (1) & 43 (0.928) \\
model 6 & 56 (0.143) & 89 (0) & 89 (0) & 52 (0.385) & 57 (0.103) & 0 (1) \\
\hline
\end{tabular}
}
\end{center}
\end{table}

\begin{figure}[h!]
\centering {
\includegraphics[width=8cm, angle=-90]{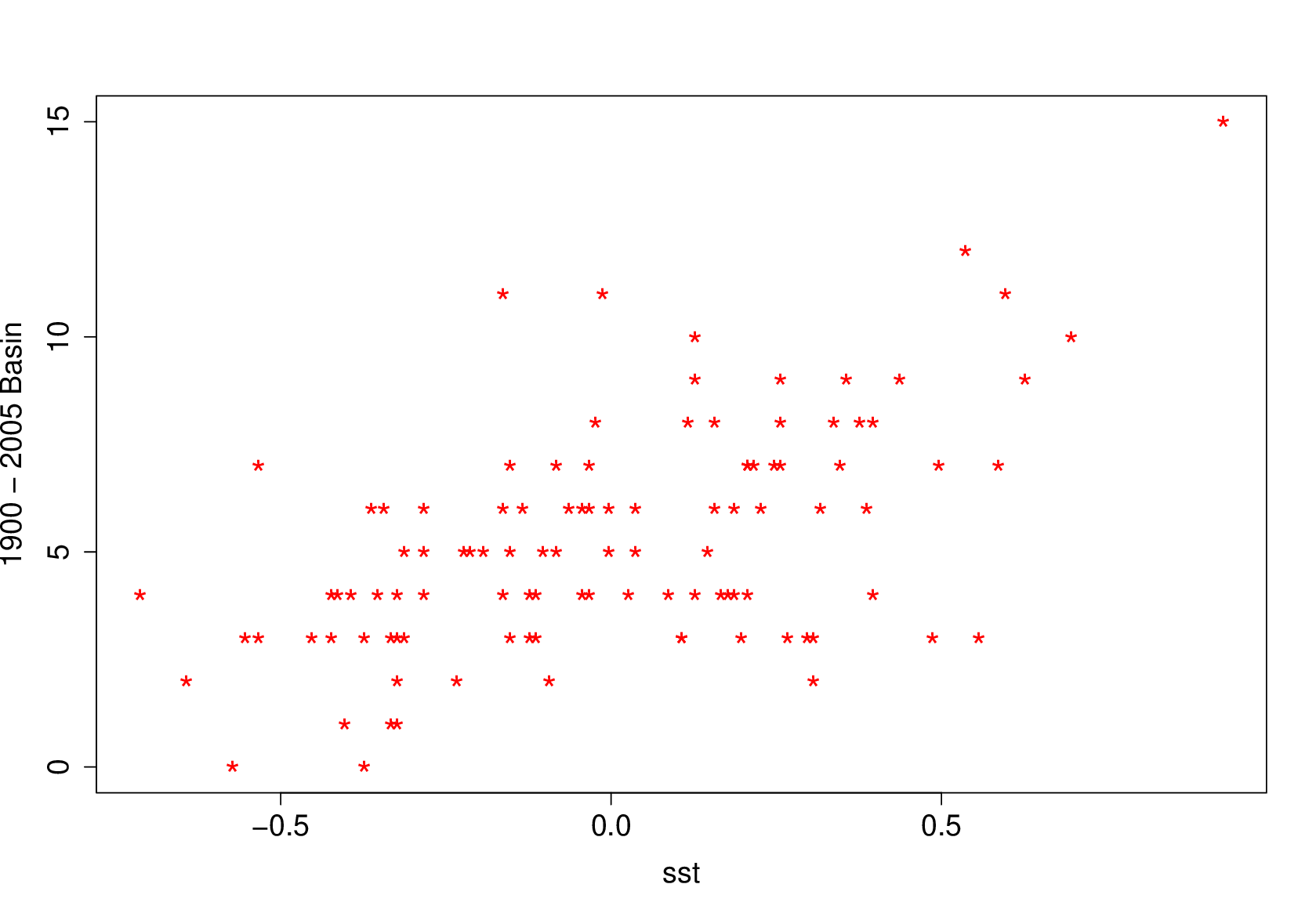}
} \caption{1900 - 2005 Basin vs SST}
\end{figure}

\begin{figure}[h!]
\centering {
\includegraphics[width=8cm, angle=-90]{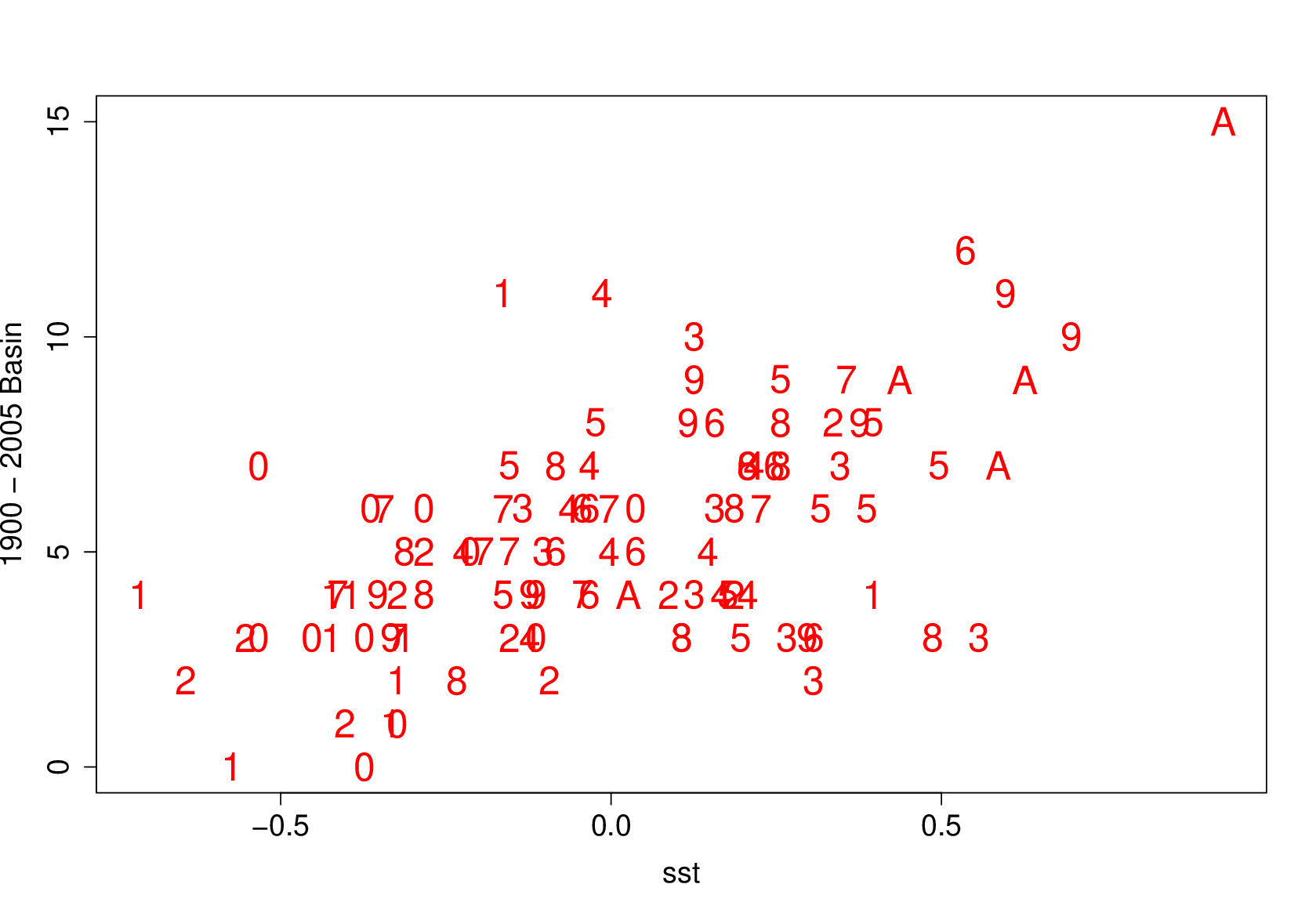}
} \caption{1900 - 2005 Basin vs SST}
\end{figure}
\clearpage

\begin{figure}[h!]
\centering {
\includegraphics[width=10cm, angle=-90]{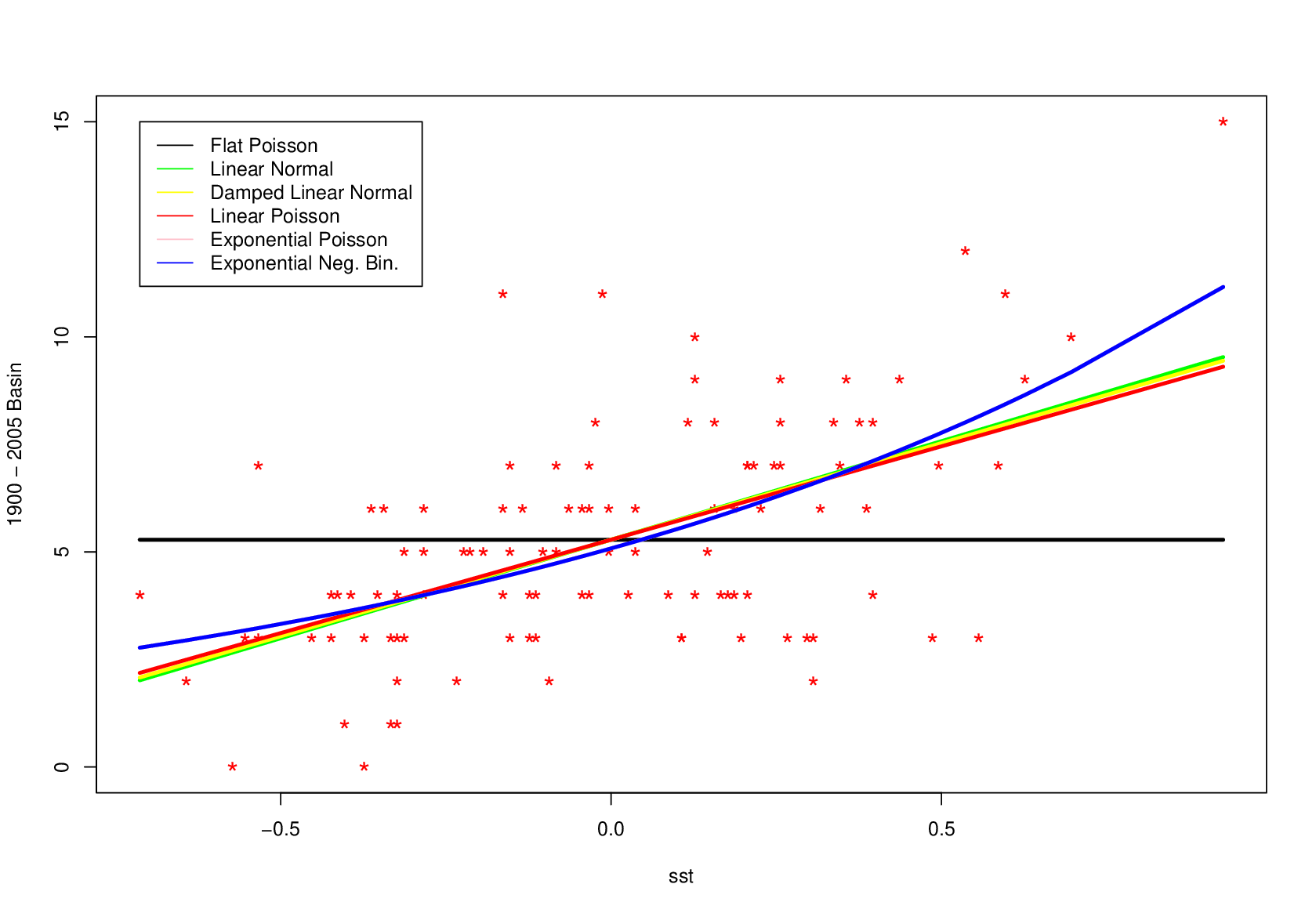}
} \caption{Fitted Lines for all Models 1900 - 2005 Basin vs SST}
\end{figure}

\newpage

\begin{table}[h!]
\begin{center}
\caption{RMSE comparison 1950 - 2005 Basin vs SST} {\small
\begin{tabular}{|c|c|c|c|c|c|c|}
\hline
 & model name & RMSE (in) & RMSE (out) & 100-100*RMSE/RMSEconst & LL (in) & LL (out) \\
\hline
model 1 & Flat Poisson & 2.607 & 2.654 & 0 & -2.332 & -2.352 \\
model 2 & Linear Normal & 2.04 & 2.13 & 35.609 & -2.507 & -2.516 \\
model 3 & Damped Linear Normal & 2.04 & 2.133 & 35.391 & -2.508 & -2.516 \\
model 4 & Linear Poisson & 2.042 & 2.122 & 36.063 & -2.137 & -2.161 \\
model 5 & Exponential Poisson & 1.981 & 2.067 & 39.335 & -2.126 & -2.149 \\
model 6 & Exponential Neg. Bin. & 1.981 & 2.002 & 43.117 & -2.126 & -2.136 \\
\hline
\end{tabular}
}
\end{center}
\end{table}
\begin{table}[h!]
\begin{center}
\caption{Model parameters incl. out of sample RMSE 1950 - 2005
Basin vs SST} {\small
\begin{tabular}{|c|c|c|c|c|c|c|c|c|}
\hline
 & $\hat{\alpha}$ & s.e. & $\hat{\beta}$ & s.e. & $k$ & cov & corr & RMSE (out of sample) \\
\hline
model 1 & 1.833 & 0.053 &  &  &  &  &  & 2.654 \\
model 2 & 6.25 & 0.278 & 5.38 & 0.92 &  & 0 & 0 & 2.13 \\
model 3 & 6.25 &  & 5.227 &  & 0.972 &  &  & 2.133 \\
model 4 & 6.25 & 0.334 & 5.021 & 1.106 &  & 0.09 & 0.243 & 2.122 \\
model 5 & 1.8 & 0.055 & 0.832 & 0.172 &  & -0.002 & -0.244 & 2.067 \\
model 6 & 1.8 & 0.055 & 0.832 & 0.172 &  & -0.001 & -0.244 & 2.002 \\
\hline
\end{tabular}
}
\end{center}
\end{table}
\begin{table}[h!]
\begin{center}
\caption{Winning count for particular model 1950 - 2005 Basin vs
SST} {\small
\begin{tabular}{|c|c|c|c|c|c|c|}
\hline
 & model 1 & model 2 & model 3 & model 4 & model 5 & model 6 \\
\hline
model 1 & 0 (1) & 38 (0.978) & 38 (0.978) & 38 (0.978) & 39 (0.959) & 39 (0.959) \\
model 2 & 62 (0.041) & 0 (1) & 54 (0.344) & 54 (0.344) & 48 (0.656) & 48 (0.656) \\
model 3 & 62 (0.041) & 46 (0.748) & 0 (1) & 52 (0.447) & 52 (0.447) & 48 (0.656) \\
model 4 & 62 (0.041) & 46 (0.748) & 48 (0.656) & 0 (1) & 54 (0.344) & 48 (0.656) \\
model 5 & 61 (0.07) & 52 (0.447) & 48 (0.656) & 46 (0.748) & 0 (1) & 50 (0.553) \\
model 6 & 61 (0.07) & 52 (0.447) & 52 (0.447) & 52 (0.447) & 50 (0.553) & 0 (1) \\
\hline
\end{tabular}
}
\end{center}
\end{table}
\begin{table}[h!]
\begin{center}
\caption{Winning count (LL) for particular model 1950 - 2005 Basin
vs SST} {\small
\begin{tabular}{|c|c|c|c|c|c|c|}
\hline
 & model 1 & model 2 & model 3 & model 4 & model 5 & model 6 \\
\hline
model 1 & 0 (1) & 77 (0) & 77 (0) & 38 (0.978) & 39 (0.959) & 39 (0.959) \\
model 2 & 23 (1) & 0 (1) & 54 (0.344) & 12 (1) & 12 (1) & 12 (1) \\
model 3 & 23 (1) & 46 (0.748) & 0 (1) & 12 (1) & 12 (1) & 11 (1) \\
model 4 & 62 (0.041) & 88 (0) & 88 (0) & 0 (1) & 54 (0.344) & 48 (0.656) \\
model 5 & 61 (0.07) & 88 (0) & 88 (0) & 46 (0.748) & 0 (1) & 50 (0.553) \\
model 6 & 61 (0.07) & 88 (0) & 89 (0) & 52 (0.447) & 50 (0.553) & 0 (1) \\
\hline
\end{tabular}
}
\end{center}
\end{table}

\begin{figure}[h!]
\centering {
\includegraphics[width=8cm, angle=-90]{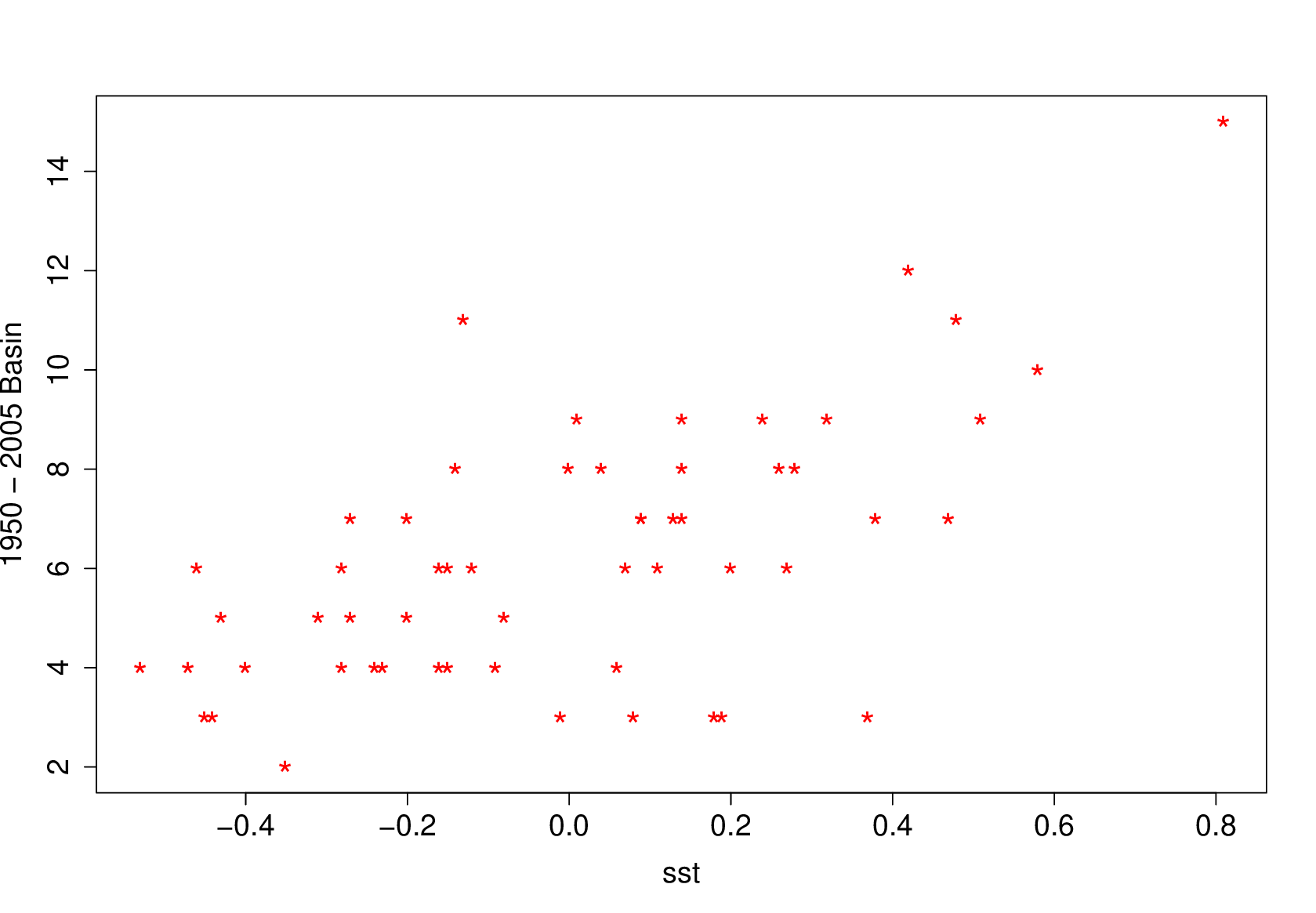}
}  \caption{1950 - 2005 Basin vs. SST}
\end{figure}

\begin{figure}[h!]
\centering {
\includegraphics[width=8cm, angle=-90]{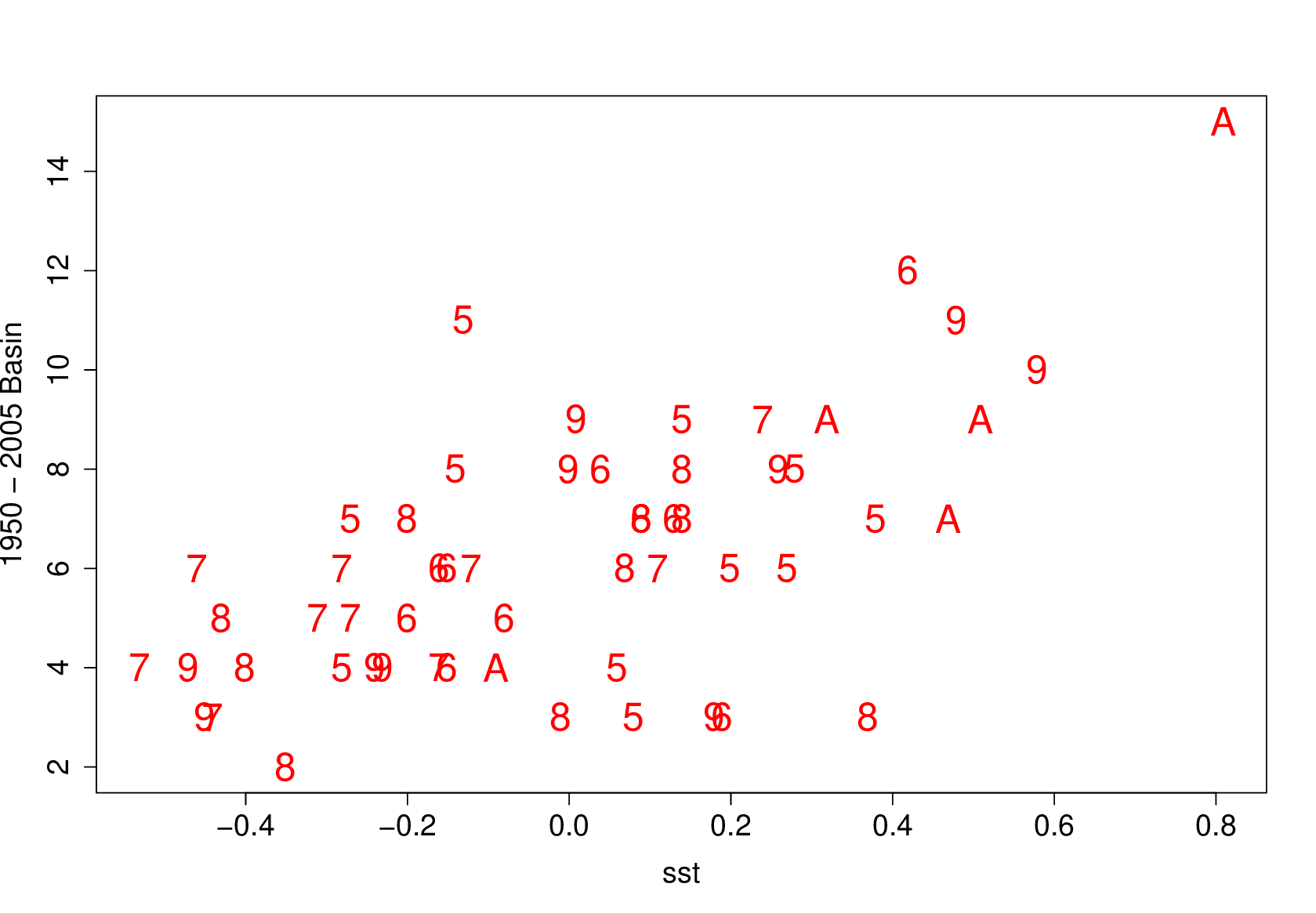}
} \caption{1950 - 2005 Basin vs. SST}
\end{figure}

\begin{figure}[h!]
\centering {
\includegraphics[width=10cm, angle=-90]{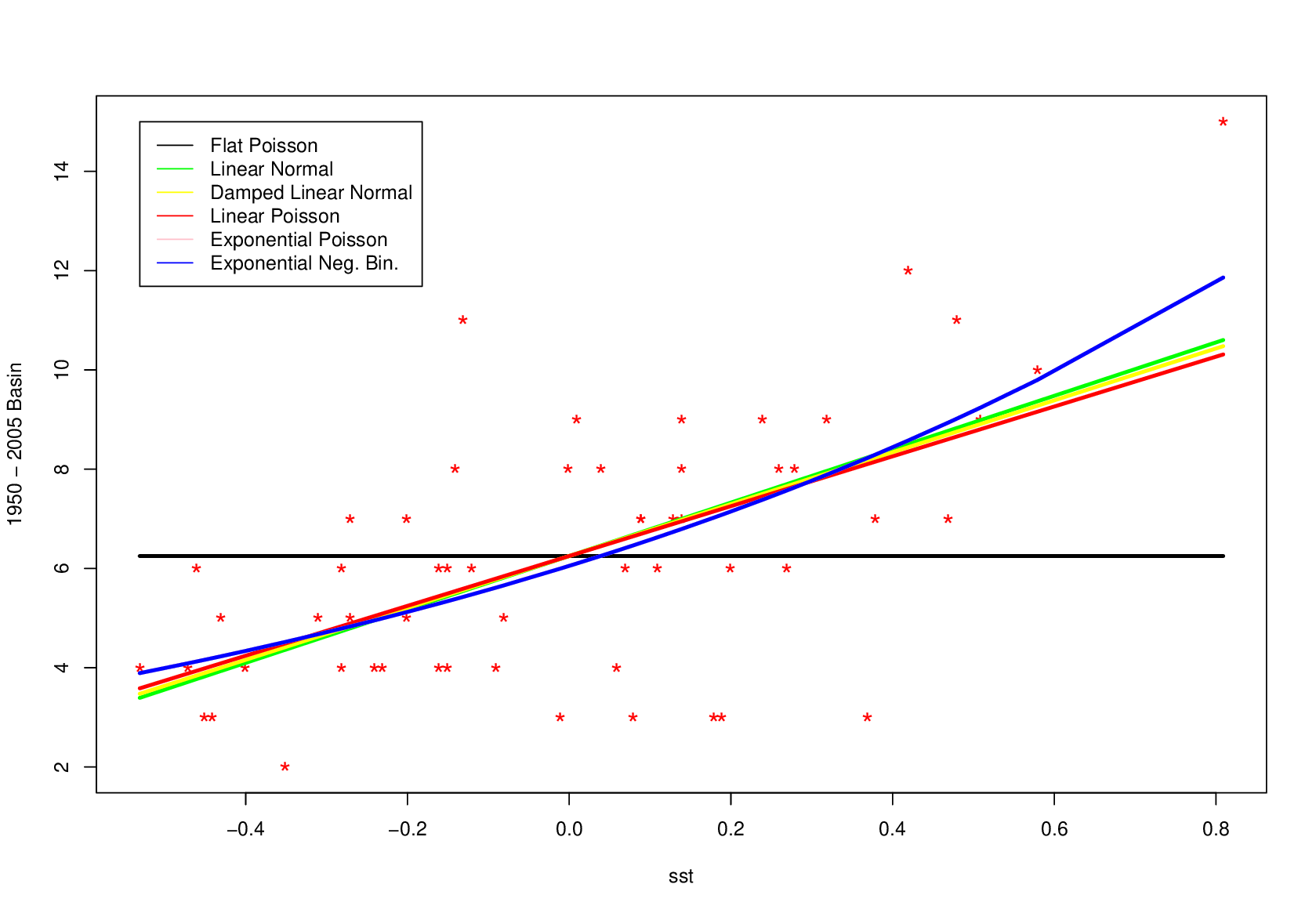}
} \caption{Fitted Lines for all Models 1950 - 2005 Basin vs SST}
\end{figure}
\clearpage
\newpage

\begin{table}[h!]
\begin{center}
\caption{RMSE comparison 1900 - 2005 Intense Basin vs SST} {\small
\begin{tabular}{|c|c|c|c|c|c|c|}
\hline
 & model name & RMSE (in) & RMSE (out) & 100-100*RMSE/RMSEconst & LL (in) & LL (out) \\
\hline
model 1 & Flat Poisson & 1.861 & 1.878 & 0 & -1.947 & -1.962 \\
model 2 & Linear Normal & 1.587 & 1.615 & 26.071 & -2.058 & -2.075 \\
model 3 & Damped Linear Normal & 1.587 & 1.616 & 26.009 & -2.059 & -2.075 \\
model 4 & Linear Poisson & 1.587 & 1.609 & 26.617 & -1.714 & -1.731 \\
model 5 & Exponential Poisson & 1.6 & 1.629 & 24.82 & -1.734 & -1.753 \\
model 6 & Exponential Neg. Bin. & 1.601 & 1.609 & 26.596 & -1.728 & -1.742 \\
\hline
\end{tabular}
}
\end{center}
\end{table}
\begin{table}[h!]
\begin{center}
\caption{Model parameters incl. out of sample RMSE 1900 - 2005
Intense Basin vs SST} {\small
\begin{tabular}{|c|c|c|c|c|c|c|c|c|}
\hline
 & $\hat{\alpha}$ & s.e. & $\hat{\beta}$ & s.e. & $k$ & cov & corr & RMSE (out of sample) \\
\hline
model 1 & 0.775 & 0.066 &  &  &  &  &  & 1.878 \\
model 2 & 2.17 & 0.156 & 2.976 & 0.476 &  & 0 & 0 & 1.615 \\
model 3 & 2.17 &  & 2.901 &  & 0.975 &  &  & 1.616 \\
model 4 & 2.17 & 0.143 & 2.947 & 0.318 &  & 0.028 & 0.612 & 1.609 \\
model 5 & 0.678 & 0.072 & 1.333 & 0.198 &  & -0.006 & -0.402 & 1.629 \\
model 6 & 0.676 & 0.077 & 1.358 & 0.218 &  & -0.006 & -0.356 & 1.609 \\
\hline
\end{tabular}
}
\end{center}
\end{table}
\begin{table}[h!]
\begin{center}
\caption{Winning count for particular model 1900 - 2005 Intense
Basin vs SST} {\small
\begin{tabular}{|c|c|c|c|c|c|c|}
\hline
 & model 1 & model 2 & model 3 & model 4 & model 5 & model 6 \\
\hline
model 1 & 0 (1) & 41 (0.98) & 41 (0.98) & 41 (0.98) & 41 (0.98) & 41 (0.98) \\
model 2 & 59 (0.032) & 0 (1) & 53 (0.314) & 47 (0.752) & 48 (0.686) & 46 (0.809) \\
model 3 & 59 (0.032) & 47 (0.752) & 0 (1) & 42 (0.951) & 49 (0.615) & 46 (0.809) \\
model 4 & 59 (0.032) & 53 (0.314) & 58 (0.072) & 0 (1) & 49 (0.615) & 47 (0.752) \\
model 5 & 59 (0.032) & 52 (0.385) & 51 (0.461) & 51 (0.461) & 0 (1) & 46 (0.809) \\
model 6 & 59 (0.032) & 54 (0.248) & 54 (0.248) & 53 (0.314) & 54 (0.248) & 0 (1) \\
\hline
\end{tabular}
}
\end{center}
\end{table}
\begin{table}[h!]
\begin{center}
\caption{Winning count (LL) for particular model 1900 - 2005
Intense Basin vs SST} {\small
\begin{tabular}{|c|c|c|c|c|c|c|}
\hline
 & model 1 & model 2 & model 3 & model 4 & model 5 & model 6 \\
\hline
model 1 & 0 (1) & 68 (0) & 68 (0) & 42 (0.968) & 41 (0.98) & 41 (0.98) \\
model 2 & 32 (1) & 0 (1) & 53 (0.314) & 15 (1) & 16 (1) & 14 (1) \\
model 3 & 32 (1) & 47 (0.752) & 0 (1) & 15 (1) & 16 (1) & 15 (1) \\
model 4 & 58 (0.049) & 85 (0) & 85 (0) & 0 (1) & 49 (0.615) & 48 (0.686) \\
model 5 & 59 (0.032) & 84 (0) & 84 (0) & 51 (0.461) & 0 (1) & 48 (0.686) \\
model 6 & 59 (0.032) & 86 (0) & 85 (0) & 52 (0.385) & 52 (0.385) & 0 (1) \\
\hline
\end{tabular}
}
\end{center}
\end{table}

\begin{figure}[h!]
\centering {
\includegraphics[width=8cm, angle=-90]{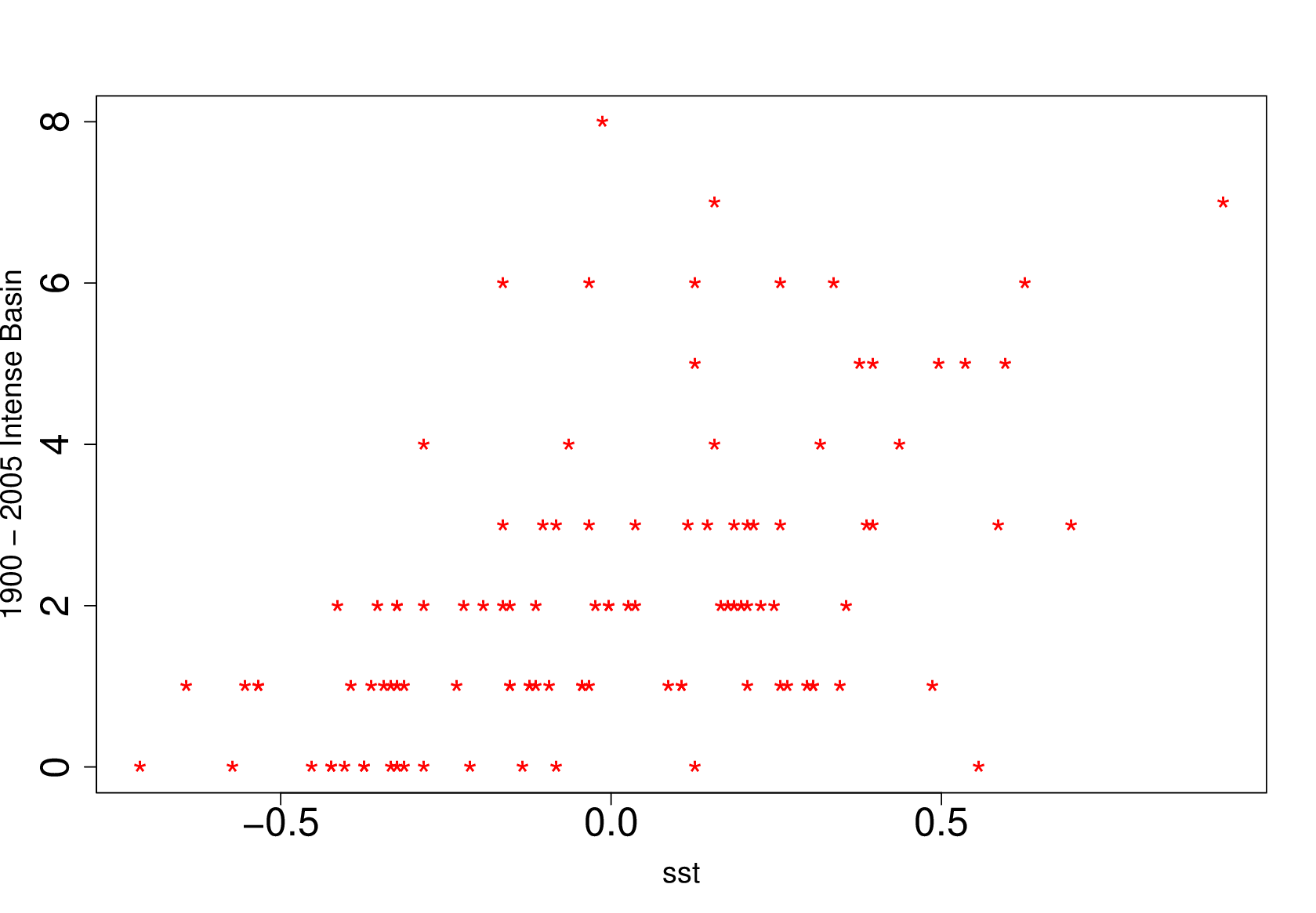}
} \caption{1900 - 2005 Intense Basin vs SST}
\end{figure}

\begin{figure}[h!]
\centering {
\includegraphics[width=8cm, angle=-90]{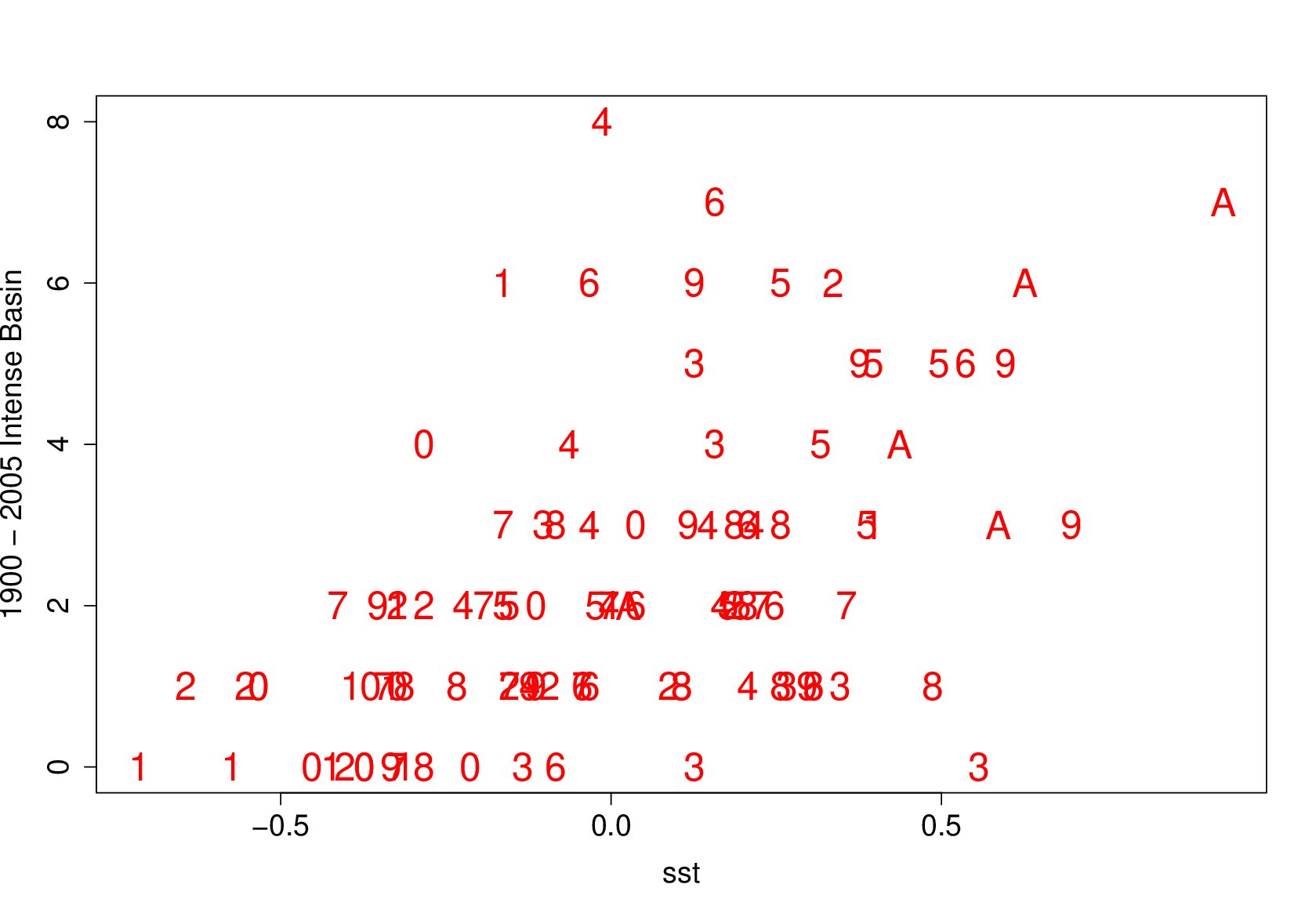}
} \caption{1900 - 2005 Intense Basin vs SST}
\end{figure}

\begin{figure}[h!]
\centering {
\includegraphics[width=10cm, angle=-90]{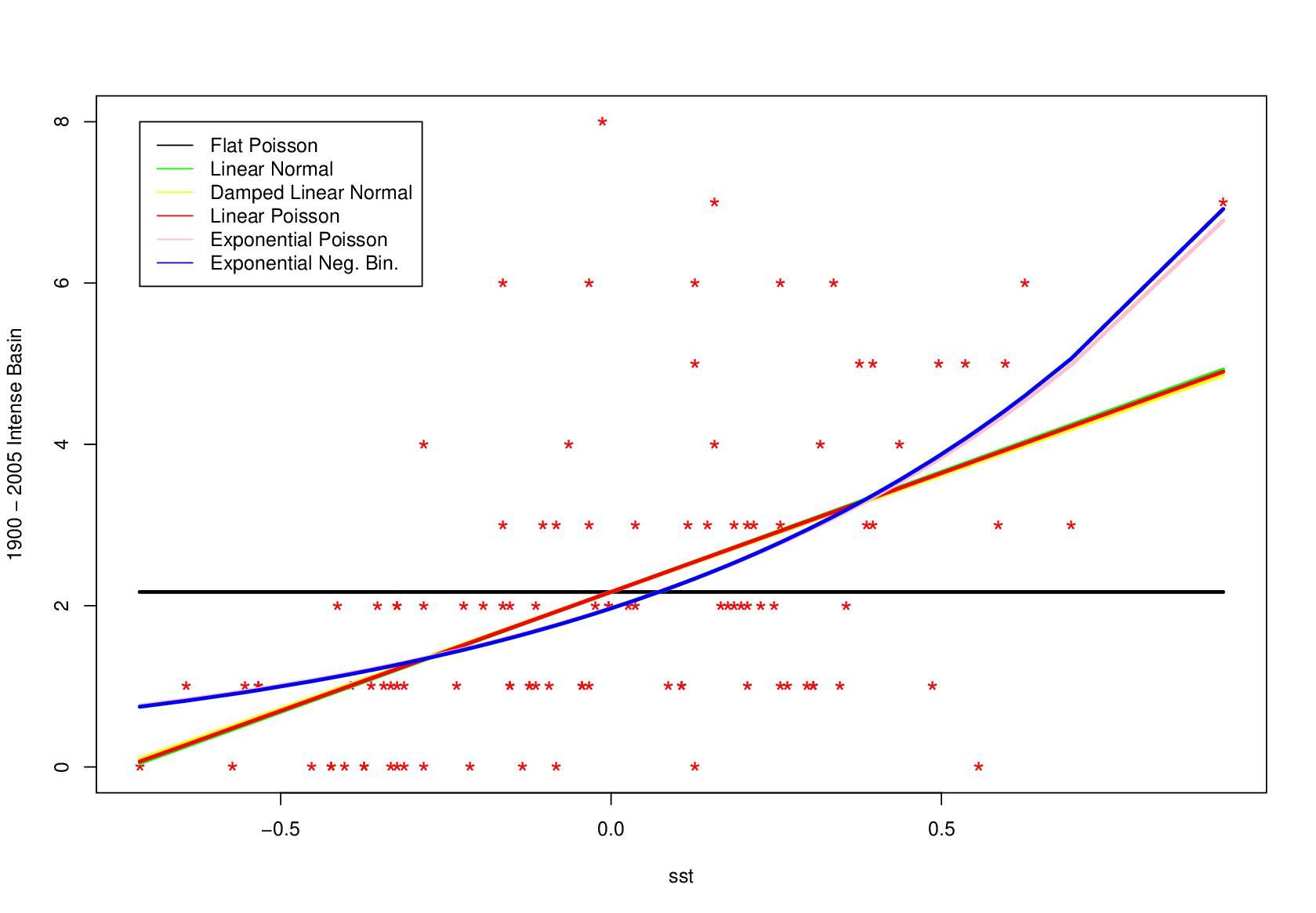}
} \caption{Fitted Lines for all Models 1900 - 2005 Intense Basin
vs SST}
\end{figure}
\clearpage
\newpage

\begin{table}[h!]
\begin{center}
\caption{RMSE comparison 1950 - 2005 Intense Basin vs SST} {\small
\begin{tabular}{|c|c|c|c|c|c|c|}
\hline
 & model name & RMSE (in) & RMSE (out) & 100-100*RMSE/RMSEconst & LL (in) & LL (out) \\
\hline
model 1 & Flat Poisson & 1.965 & 2.001 & 0 & -2.018 & -2.044 \\
model 2 & Linear Normal & 1.664 & 1.715 & 26.517 & -2.154 & -2.184 \\
model 3 & Damped Linear Normal & 1.664 & 1.718 & 26.294 & -2.154 & -2.184 \\
model 4 & Linear Poisson & 1.664 & 1.711 & 26.868 & -1.811 & -1.847 \\
model 5 & Exponential Poisson & 1.668 & 1.717 & 26.307 & -1.82 & -1.853 \\
model 6 & Exponential Neg. Bin. & 1.668 & 1.693 & 28.392 & -1.82 & -1.837 \\
\hline
\end{tabular}
}
\end{center}
\end{table}
\begin{table}[h!]
\begin{center}
\caption{Model parameters incl. out of sample RMSE 1950 - 2005
Intense Basin vs SST} {\small
\begin{tabular}{|c|c|c|c|c|c|c|c|c|}
\hline
 & $\hat{\alpha}$ & s.e. & $\hat{\beta}$ & s.e. & $k$ & cov & corr & RMSE (out of sample) \\
\hline
model 1 & 0.985 & 0.082 &  &  &  &  &  & 2.001 \\
model 2 & 2.679 & 0.226 & 3.466 & 0.75 &  & 0 & 0 & 1.715 \\
model 3 & 2.679 &  & 3.311 &  & 0.955 &  &  & 1.718 \\
model 4 & 2.679 & 0.219 & 3.435 & 0.692 &  & 0.061 & 0.405 & 1.711 \\
model 5 & 0.914 & 0.087 & 1.235 & 0.26 &  & -0.008 & -0.351 & 1.717 \\
model 6 & 0.913 & 0.089 & 1.238 & 0.266 &  & -0.009 & -0.34 & 1.693 \\
\hline
\end{tabular}
}
\end{center}
\end{table}
\begin{table}[h!]
\begin{center}
\caption{Winning count for particular model 1950 - 2005 Intense
Basin vs SST} {\small
\begin{tabular}{|c|c|c|c|c|c|c|}
\hline
 & model 1 & model 2 & model 3 & model 4 & model 5 & model 6 \\
\hline
model 1 & 0 (1) & 36 (0.989) & 36 (0.989) & 36 (0.989) & 38 (0.978) & 36 (0.989) \\
model 2 & 64 (0.022) & 0 (1) & 57 (0.175) & 54 (0.344) & 43 (0.886) & 46 (0.748) \\
model 3 & 64 (0.022) & 43 (0.886) & 0 (1) & 41 (0.93) & 41 (0.93) & 43 (0.886) \\
model 4 & 64 (0.022) & 46 (0.748) & 59 (0.114) & 0 (1) & 43 (0.886) & 45 (0.825) \\
model 5 & 62 (0.041) & 57 (0.175) & 59 (0.114) & 57 (0.175) & 0 (1) & 45 (0.825) \\
model 6 & 64 (0.022) & 54 (0.344) & 57 (0.175) & 55 (0.252) & 55 (0.252) & 0 (1) \\
\hline
\end{tabular}
}
\end{center}
\end{table}
\begin{table}[h!]
\begin{center}
\caption{Winning count (LL) for particular model 1950 - 2005
Intense Basin vs SST} {\small
\begin{tabular}{|c|c|c|c|c|c|c|}
\hline
 & model 1 & model 2 & model 3 & model 4 & model 5 & model 6 \\
\hline
model 1 & 0 (1) & 71 (0.001) & 71 (0.001) & 36 (0.989) & 39 (0.959) & 36 (0.989) \\
model 2 & 29 (1) & 0 (1) & 57 (0.175) & 12 (1) & 11 (1) & 12 (1) \\
model 3 & 29 (1) & 43 (0.886) & 0 (1) & 12 (1) & 11 (1) & 12 (1) \\
model 4 & 64 (0.022) & 88 (0) & 88 (0) & 0 (1) & 43 (0.886) & 45 (0.825) \\
model 5 & 61 (0.07) & 89 (0) & 89 (0) & 57 (0.175) & 0 (1) & 43 (0.886) \\
model 6 & 64 (0.022) & 88 (0) & 88 (0) & 55 (0.252) & 57 (0.175) & 0 (1) \\
\hline
\end{tabular}
}
\end{center}
\end{table}

\begin{figure}[h!]
\centering {
\includegraphics[width=8cm, angle=-90]{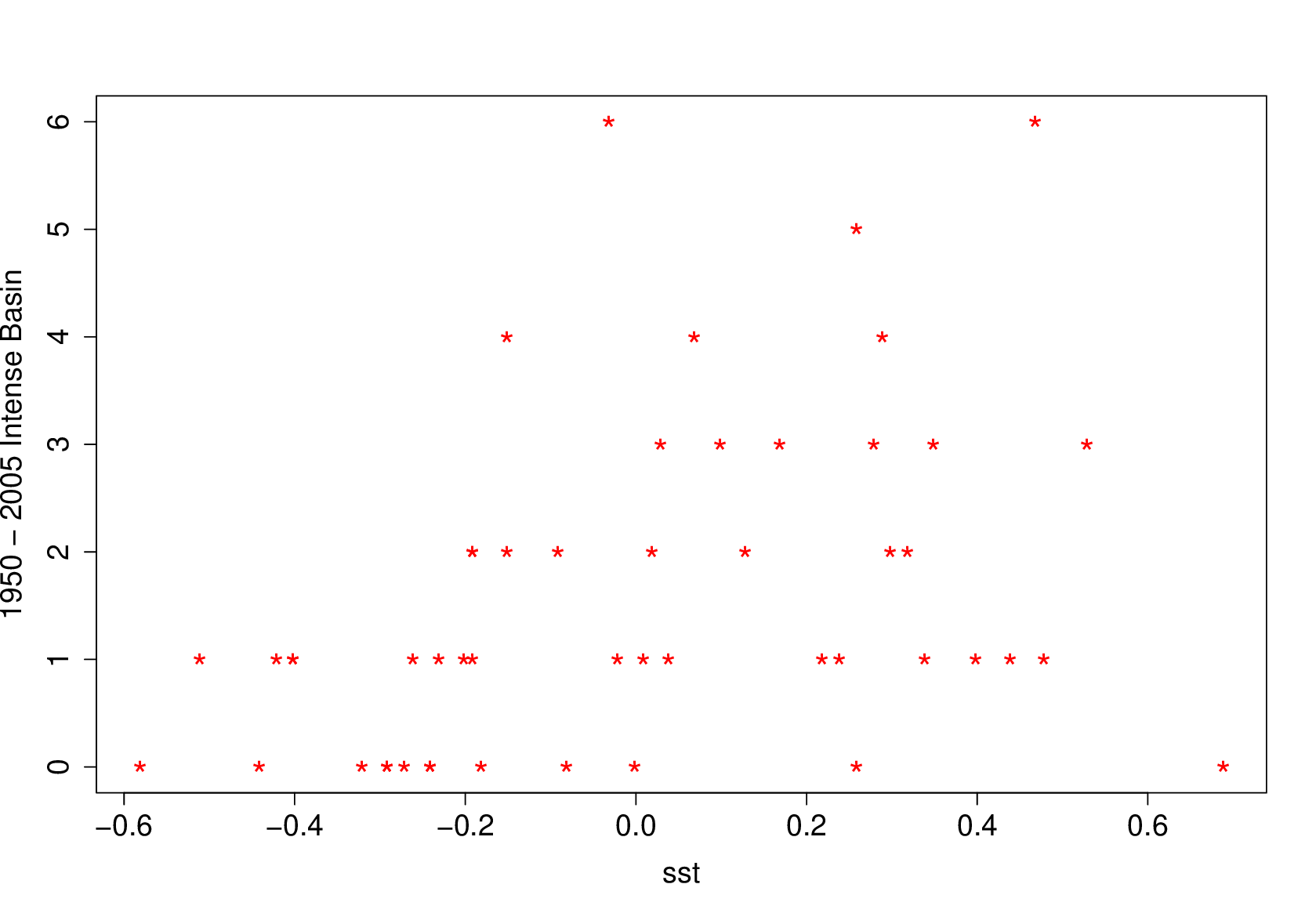}
} \caption{1950 - 2005 Intense Basin vs SST}
\end{figure}

\begin{figure}[h!]
\centering {
\includegraphics[width=8cm, angle=-90]{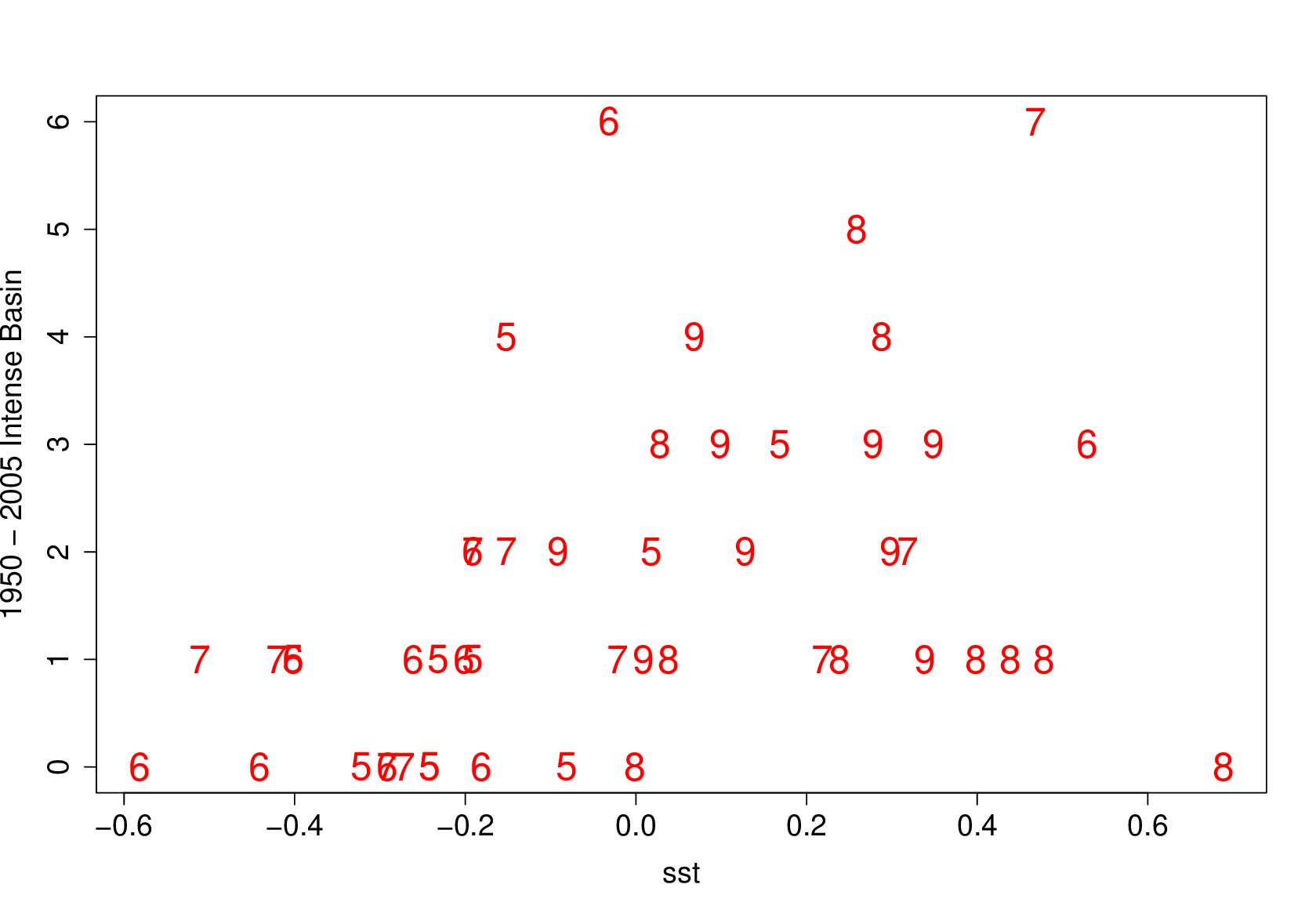}
} \caption{1950 - 2005 Intense Basin vs SST}
\end{figure}

\begin{figure}[h!]
\centering {
\includegraphics[width=10cm, angle=-90]{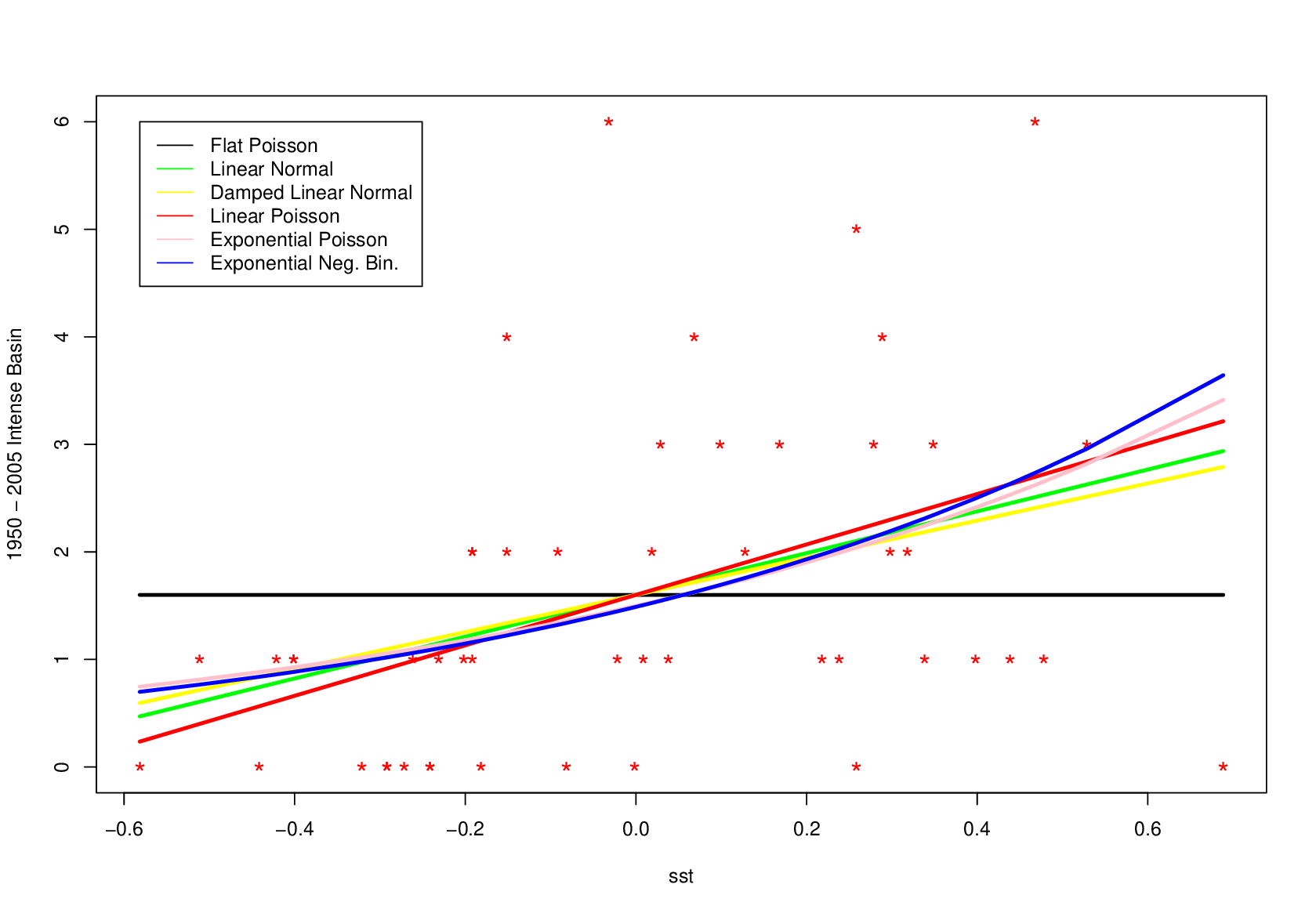}
} \caption{Fitted Lines for all Models 1950 - 2005 Intense Basin
vs SST}
\end{figure}

\newpage
 \setlength{\hoffset}{-1.5in}\setlength           {\oddsidemargin}                                                 {2.0cm}
 \setlength{\textwidth}{12cm} \setlength           {\voffset}                                                       {-1in}
 \setlength{\topmargin}{1cm} \setlength            {\textheight}{25cm}
 \setlength{\unitlength}{1cm} \setlength           {\parindent}{2cm}
 \begin{table}[ht]
   \centering
 \begin{tabular}{|c|c|c|c|c|c|c|c|c|}
  \hline
  1 & 2 & 3 & 4 & 5 & 6 & 7 & 8 & 9 \\
  \hline
 No. & SST Model(Window) & SST2HU Model  &              2006 Mean &                                                       2007 Mean & 2008 Mean & 2009 Mean & 2010 Mean                   & Mean \\
  \hline
         1& FL(8) & Linear Poisson & 8.040& 8.040& 8.040& 8.040& 8.040& 8.040\\
         2& DLT & Linear Poisson & 8.353& 8.441& 8.529& 8.616& 8.704& 8.529\\
         3& LT(22) & Linear Poisson & 8.667& 8.842& 9.018& 9.193& 9.368& 9.018\\
         4& FL(8) & Exp. Poisson & 8.368& 8.381& 8.383& 8.387& 8.388& 8.381\\
         5& DLT & Exp. Poisson & 8.807& 8.952& 9.092& 9.240& 9.379& 9.094\\
         6& LT(22) & Exp. Poisson & 9.295& 9.596& 9.897& 10.217& 10.524& 9.906\\
 \hline
 \end{tabular}
 \caption{
Predictions of mean basin hurricane numbers.
 }\label{b01}
 \end{table}

 \begin{table}[h!]
   \centering
 \begin{tabular}{|c|c|c|c|c|c|c|c|c|}
  \hline
  1 & 2 & 3 & 4 & 5 & 6 & 7 & 8 & 9 \\
  \hline
 No. & SST Model(Window) & SST2HU Model  &              2006 Var &                                                        2007 Var & 2008 Var & 2009 Var & 2010 Var                       & Mean Var \\
  \hline
         1& FL(8) & Linear Poisson & 10.065& 10.185& 10.203& 10.235& 10.239& 10.185\\
         2& DLT & Linear Poisson & 10.320& 10.539& 10.696& 10.905& 11.025& 10.697\\
         3& LT(22) & Linear Poisson & 10.775& 11.160& 11.464& 11.843& 12.059& 11.460\\
         4& FL(8) & Exp. Poisson & 12.372& 12.643& 12.684& 12.757& 12.767& 12.645\\
         5& DLT & Exp. Poisson & 13.110& 13.704& 14.161& 14.778& 15.169& 14.184\\
         6& LT(22) & Exp. Poisson & 14.442& 15.646& 16.701& 18.094& 19.017& 16.780\\
 \hline
 \end{tabular}
 \caption{
Prediction of the variance of basin hurricane numbers.
 }\label{b02}
 \end{table}

 \begin{table}[h!]
   \centering
 \begin{tabular}{|c|c|c|c|c|c|c|c|}
  \hline
  1 & 2 & 3 & 4 & 5 & 6 & 7 & 8 \\
  \hline
 No. & SST Model & SST2HU Model                       & 2006(T1,T2)                                                     & 2007(T1,T2)                                                     & 2008(T1,T2)                                                     & 2009(T1,T2)                                                     & 2010(T1,T2)                                                     \\
  \hline
         1& FL(8)  & Linear Poisson & 20, 80&21, 79&21, 79&21, 79&21, 79\\
         2& DLT    & Linear Poisson & 19, 81&20, 80&20, 80&21, 79&21, 79\\
         3& LT(22) & Linear Poisson & 20, 80&21, 79&21, 79&22, 78&22, 78\\
         4& FL(8)  & Exp. Poisson   & 32, 68&34, 66&34, 66&34, 66&34, 66\\
         5& DLT    & Exp. Poisson   & 33, 67&35, 65&36, 64&37, 63&38, 62\\
         6& LT(22) & Exp. Poisson   & 36, 64&39, 61&41, 59&44, 56&45, 55\\
 \hline
 \end{tabular}
 \caption{
Breakdown of the variance in table~\ref{b02} into variance driven by SST prediction uncertainty and uncertainty in the SST-to-hurricane numbers regression model.
 }\label{b03}
 \end{table}

 \begin{table}[h!]
   \centering
 \begin{tabular}{|c|c|c|c|c|c|c|c|c|}
  \hline
  1 & 2 & 3 & 4 & 5 & 6 & 7 & 8 & 9 \\
  \hline
 No. & SST Model(Win) & SST2HU Model & 2006 V/M &       2007 V/M & 2008 V/M & 2009 V/M & 2010 V/M                       & Mean V/M \\
  \hline
         1& FL(8) & Linear Poisson & 1.252& 1.267& 1.269& 1.273& 1.274& 1.267\\
         2& DLT & Linear Poisson & 1.235& 1.249& 1.254& 1.266& 1.267& 1.254\\
         3& LT(22) & Linear Poisson & 1.243& 1.262& 1.271& 1.288& 1.287& 1.271\\
         4& FL(8) & Exp. Poisson & 1.479& 1.508& 1.513& 1.521& 1.522& 1.509\\
         5& DLT & Exp. Poisson & 1.489& 1.531& 1.558& 1.599& 1.617& 1.560\\
         6& LT(22) & Exp. Poisson & 1.554& 1.630& 1.688& 1.771& 1.807& 1.694\\
 \hline
 \end{tabular}
 \caption{
Ratio of the variance to the mean.
 }\label{b04}
 \end{table}

 \begin{table}[h!]
   \centering
 \begin{tabular}{|c|c|c|c|c|c|c|c|c|}
  \hline
  1 & 2 & 3 & 4 & 5 & 6 & 7 & 8 & 9 \\
  \hline
 No. & SST Model & S2B &                                2006 RMSE & 2007 RMSE & 2008 RMSE & 2009 RMSE & 2010 RMSE &       Mean\\
  \hline
         1& FL(8) & Linear Poisson & 0.764& 0.774& 0.776& 0.778& 0.779& 0.774\\
         2& DLT & Linear Poisson & 0.975& 1.044& 1.113& 1.189& 1.269& 1.118\\
         3& LT(22) & Linear Poisson & 1.199& 1.331& 1.475& 1.629& 1.796& 1.486\\
         4& FL(8) & Exp. Poisson & 1.247& 1.266& 1.268& 1.273& 1.274& 1.266\\
         5& DLT & Exp. Poisson & 1.612& 1.744& 1.878& 2.029& 2.187& 1.890\\
         6& LT(22) & Exp. Poisson & 2.051& 2.334& 2.650& 3.011& 3.405& 2.690\\
 \hline
 \end{tabular}
 \caption{
Standard errors on the predicted means.
 }\label{b05}
 \end{table}
 \clearpage

 \begin{table}[h!]
   \centering
 \begin{tabular}{|c|c|c|c|c|c|c|c|c|c|}
  \hline
  1 & 2 & 3 & 4 & 5 & 6 & 7 & 8 & 9 & 10 \\
  \hline
 No. & SST Model & S2B Model & B2L Model &              2006 Mean &                                                       2007 Mean & 2008 Mean & 2009 Mean & 2010 Mean                   & Mean \\
  \hline
  1& FL(8) & LinPois & PoisConProp & 2.044& 2.044& 2.044& 2.044& 2.044& 2.044\\
  2& DLT & LinPois & PoisConProp & 2.124& 2.146& 2.169& 2.191& 2.213& 2.169\\
  3& LT(22) & LinPois & PoisConProp & 2.204& 2.248& 2.293& 2.338& 2.382& 2.293\\
  4& FL(8) & ExpPois & PoisConProp & 2.128& 2.131& 2.132& 2.133& 2.133& 2.131\\
  5& DLT & ExpPois & PoisConProp & 2.239& 2.276& 2.312& 2.350& 2.385& 2.312\\
  6& LT(22) & ExpPois & PoisConProp & 2.364& 2.440& 2.517& 2.598& 2.676& 2.519\\
 \hline
 \end{tabular}
 \caption{
Predictions of mean landfalling hurricane numbers.
 }\label{b06}
 \end{table}

 \begin{table}[h!]
   \centering
 \begin{tabular}{|c|c|c|c|c|c|c|c|c|c|}
  \hline
  1 & 2 & 3 & 4 & 5 & 6 & 7 & 8 & 9 & 10 \\
  \hline
 No. & SST Model & S2B Model & B2L Model &              2006 Var &                                                        2007 Var & 2008 Var & 2009 Var & 2010 Var                       & Mean Var \\
  \hline
  1& FL(8)  & LinPois & PoisConProp & 2.695& 2.703& 2.704& 2.706& 2.706& 2.703\\
  2& DLT    & LinPois & PoisConProp & 2.791& 2.828& 2.860& 2.896& 2.926& 2.860\\
  3& LT(22) & LinPois & PoisConProp & 2.901& 2.970& 3.034& 3.103& 3.162& 3.034\\
  4& FL(8)  & ExpPois & PoisConProp & 2.928& 2.949& 2.952& 2.958& 2.958& 2.949\\
  5& DLT    & ExpPois & PoisConProp & 3.087& 3.162& 3.228& 3.305& 3.366& 3.230\\
  6& LT(22) & ExpPois & PoisConProp & 3.297& 3.452& 3.597& 3.768& 3.906& 3.604\\
 \hline
 \end{tabular}
 \caption{
Prediction of the variance of landfalling hurricane numbers.
 }\label{b07}
 \end{table}

 \begin{table}[h!]
   \centering
 \begin{tabular}{|c|c|c|c|c|c|c|c|c|}
  \hline
  1 & 2 & 3 & 4 & 5 & 6 & 7 & 8 & 9\\
  \hline
 No. & SST Model & SST2HU Model                       & 2006(T1,T2)                                                     & 2007(T1,T2)                                                     & 2008(T1,T2)                                                     & 2009(T1,T2)                                                     & 2010(T1,T2)                                                     \\
  \hline
  1& FL(8)  & LinPois & PoisConProp & 24, 76&24, 76&24, 76&24, 76&24, 76\\
  2& DLT    & LinPois & PoisConProp & 24, 76&24, 76&24, 76&24, 76&24, 76\\
  3& LT(22) & LinPois & PoisConProp & 24, 76&24, 76&24, 76&25, 75&25, 75\\
  4& FL(8)  & ExpPois & PoisConProp & 27, 73&28, 72&28, 72&28, 72&28, 72\\
  5& DLT    & ExpPois & PoisConProp & 27, 73&28, 72&28, 72&29, 71&29, 71\\
  6& LT(22) & ExpPois & PoisConProp & 28, 72&29, 71&30, 70&31, 69&31, 69\\
 \hline
 \end{tabular}
 \caption{
Breakdown of the variance in table~\ref{b07} into variance driven by SST prediction uncertainty and uncertainty in the SST-to-hurricane numbers regression model.
 }\label{b08}
 \end{table}

 \begin{table}[h!]
   \centering
 \begin{tabular}{|c|c|c|c|c|c|c|c|c|c|}
  \hline
  1 & 2 & 3 & 4 & 5 & 6 & 7 & 8 & 9 & 10 \\
  \hline
 No. & SST Model & S2B Model & B2L Model               & 2006 V/M &  2007 V/M & 2008 V/M & 2009 V/M & 2010 V/M          & Mean V/M \\
  \hline
  1& FL(8) & LinPois & PoisConProp & 1.318& 1.322& 1.323& 1.324& 1.324& 1.322\\
  2& DLT & LinPois & PoisConProp & 1.314& 1.317& 1.319& 1.322& 1.322& 1.319\\
  3& LT(22) & LinPois & PoisConProp & 1.316& 1.321& 1.323& 1.328& 1.327& 1.323\\
  4& FL(8) & ExpPois & PoisConProp & 1.376& 1.384& 1.385& 1.387& 1.387& 1.384\\
  5& DLT & ExpPois & PoisConProp & 1.379& 1.389& 1.396& 1.407& 1.411& 1.397\\
  6& LT(22) & ExpPois & PoisConProp & 1.395& 1.415& 1.429& 1.450& 1.459& 1.431\\
 \hline
 \end{tabular}
 \caption{
Ratio of the variance to the mean.
 }\label{b09}
 \end{table}

 \begin{table}[h!]
   \centering
 \begin{tabular}{|c|c|c|c|c|c|c|c|c|c|}
  \hline
  1 & 2 & 3 & 4 & 5 & 6 & 7 & 8 & 9 & 10 \\
  \hline
 No. & SST Model & S2B & B2L &                          2006 RMSE & 2007 RMSE & 2008 RMSE & 2009 RMSE & 2010 RMSE &       Mean\\
  \hline
  1& FL(8) & LinPois & PoisConProp & 0.562& 0.563& 0.563& 0.563& 0.563& 0.563\\
  2& DLT & LinPois & PoisConProp & 0.601& 0.614& 0.627& 0.641& 0.656& 0.628\\
  3& LT(22) & LinPois & PoisConProp & 0.645& 0.672& 0.700& 0.732& 0.766& 0.703\\
  4& FL(8) & ExpPois & PoisConProp & 0.634& 0.637& 0.638& 0.638& 0.639& 0.637\\
  5& DLT & ExpPois & PoisConProp & 0.708& 0.736& 0.764& 0.796& 0.829& 0.767\\
  6& LT(22) & ExpPois & PoisConProp & 0.802& 0.865& 0.936& 1.018& 1.107& 0.946\\
 \hline
 \end{tabular}
 \caption{
Standard errors on the predicted means.
 }\label{b10}
 \end{table}
 \clearpage

 \begin{table}[h!]
   \centering
 \begin{tabular}{|c|c|c|c|c|c|c|c|c|}
  \hline
  1 & 2 & 3 & 4 & 5 & 6 & 7 & 8 & 9 \\
  \hline
 No. & SST Model(Window) & SST2HU Model  &              2006 Mean &                                                       2007 Mean & 2008 Mean & 2009 Mean & 2010 Mean                   & Mean \\
  \hline
         1& FL(8) & Linear Poisson & 3.903& 3.903& 3.903& 3.903& 3.903& 3.903\\
         2& DLT & Linear Poisson & 4.118& 4.178& 4.238& 4.298& 4.358& 4.238\\
         3& LT(22) & Linear Poisson & 4.333& 4.453& 4.572& 4.692& 4.812& 4.572\\
         4& FL(8) & Exp. Poisson & 4.118& 4.133& 4.136& 4.140& 4.140& 4.133\\
         5& DLT & Exp. Poisson & 4.441& 4.556& 4.665& 4.784& 4.893& 4.668\\
         6& LT(22) & Exp. Poisson & 4.818& 5.062& 5.306& 5.574& 5.826& 5.317\\
 \hline
 \end{tabular}
 \caption{
Predictions of mean basin intense hurricane numbers.
 }\label{b11}
 \end{table}

 \begin{table}[h!]
   \centering
 \begin{tabular}{|c|c|c|c|c|c|c|c|c|}
  \hline
  1 & 2 & 3 & 4 & 5 & 6 & 7 & 8 & 9 \\
  \hline
 No. & SST Model(Window) & SST2HU Model  &              2006 Var &                                                        2007 Var & 2008 Var & 2009 Var & 2010 Var                       & Mean Var \\
  \hline
         1& FL(8) & Linear Poisson & 4.851& 4.907& 4.916& 4.931& 4.933& 4.908\\
         2& DLT & Linear Poisson & 5.038& 5.160& 5.252& 5.369& 5.444& 5.253\\
         3& LT(22) & Linear Poisson & 5.319& 5.537& 5.717& 5.932& 6.072& 5.716\\
         4& FL(8) & Exp. Poisson & 6.330& 6.501& 6.527& 6.573& 6.580& 6.502\\
         5& DLT & Exp. Poisson & 6.933& 7.365& 7.714& 8.183& 8.503& 7.740\\
         6& LT(22) & Exp. Poisson & 7.974& 8.918& 9.795& 10.975& 11.828& 9.898\\
 \hline
 \end{tabular}
 \caption{
Prediction of the variance of basin intense hurricane numbers.
 }\label{b12}
 \end{table}

 \begin{table}[h!]
   \centering
 \begin{tabular}{|c|c|c|c|c|c|c|c|}
  \hline
  1 & 2 & 3 & 4 & 5 & 6 & 7 & 8 \\
  \hline
 No. & SST Model & SST2HU Model                       & 2006(T1,T2)                                                     & 2007(T1,T2)                                                     & 2008(T1,T2)                                                     & 2009(T1,T2)                                                     & 2010(T1,T2)                                                     \\
  \hline
         1& FL(8)  & Linear Poisson & 20, 80&20, 80&21, 79&21, 79&21, 79\\
         2& DLT    & Linear Poisson & 18, 82&19, 81&19, 81&20, 80&20, 80\\
         3& LT(22) & Linear Poisson & 19, 81&20, 80&20, 80&21, 79&21, 79\\
         4& FL(8)  & Exp. Poisson   & 35, 65&36, 64&37, 63&37, 63&37, 63\\
         5& DLT    & Exp. Poisson   & 36, 64&38, 62&40, 60&42, 58&42, 58\\
         6& LT(22) & Exp. Poisson   & 40, 60&43, 57&46, 54&49, 51&51, 49\\
 \hline
 \end{tabular}
 \caption{
Breakdown of the variance in table~\ref{b12} into variance driven by SST prediction uncertainty and uncertainty in the SST-to-hurricane numbers regression model.
 }\label{b13}
 \end{table}

 \begin{table}[h!]
   \centering
 \begin{tabular}{|c|c|c|c|c|c|c|c|c|}
  \hline
  1 & 2 & 3 & 4 & 5 & 6 & 7 & 8 & 9 \\
  \hline
 No. & SST Model(Win) & SST2HU Model & 2006 V/M &       2007 V/M & 2008 V/M & 2009 V/M & 2010 V/M                       & Mean V/M \\
  \hline
         1& FL(8) & Linear Poisson & 1.243& 1.257& 1.259& 1.263& 1.264& 1.257\\
         2& DLT & Linear Poisson & 1.224& 1.235& 1.239& 1.249& 1.249& 1.239\\
         3& LT(22) & Linear Poisson & 1.228& 1.244& 1.250& 1.264& 1.262& 1.250\\
         4& FL(8) & Exp. Poisson & 1.537& 1.573& 1.578& 1.588& 1.589& 1.573\\
         5& DLT & Exp. Poisson & 1.561& 1.617& 1.654& 1.711& 1.738& 1.658\\
         6& LT(22) & Exp. Poisson & 1.655& 1.762& 1.846& 1.969& 2.030& 1.862\\
 \hline
 \end{tabular}
 \caption{
Ratio of the variance to the mean.
 }\label{b14}
 \end{table}

 \begin{table}[h!]
   \centering
 \begin{tabular}{|c|c|c|c|c|c|c|c|c|}
  \hline
  1 & 2 & 3 & 4 & 5 & 6 & 7 & 8 & 9 \\
  \hline
 No. & SST Model & S2B &                                2006 RMSE & 2007 RMSE & 2008 RMSE & 2009 RMSE & 2010 RMSE &       Mean\\
  \hline
         1& FL(8) & Linear Poisson & 0.520& 0.527& 0.528& 0.530& 0.530& 0.527\\
         2& DLT & Linear Poisson & 0.663& 0.710& 0.757& 0.809& 0.863& 0.760\\
         3& LT(22) & Linear Poisson & 0.816& 0.905& 1.003& 1.108& 1.222& 1.011\\
         4& FL(8) & Exp. Poisson & 0.859& 0.872& 0.874& 0.878& 0.878& 0.872\\
         5& DLT & Exp. Poisson & 1.097& 1.188& 1.280& 1.386& 1.496& 1.289\\
         6& LT(22) & Exp. Poisson & 1.401& 1.608& 1.839& 2.110& 2.405& 1.873\\
 \hline
 \end{tabular}
 \caption{
Standard errors on the predicted means.
 }\label{b15}
 \end{table}
 \clearpage

 \begin{table}[h!]
   \centering
 \begin{tabular}{|c|c|c|c|c|c|c|c|c|c|}
  \hline
  1 & 2 & 3 & 4 & 5 & 6 & 7 & 8 & 9 & 10 \\
  \hline
 No. & SST Model & S2B Model & B2L Model &              2006 Mean &                                                       2007 Mean & 2008 Mean & 2009 Mean & 2010 Mean                   & Mean \\
  \hline
  1& FL(8) & LinPois & PoisConProp & 0.937& 0.937& 0.937& 0.937& 0.937& 0.937\\
  2& DLT & LinPois & PoisConProp & 0.988& 1.003& 1.017& 1.031& 1.046& 1.017\\
  3& LT(22) & LinPois & PoisConProp & 1.040& 1.069& 1.097& 1.126& 1.155& 1.097\\
  4& FL(8) & ExpPois & PoisConProp & 0.988& 0.992& 0.993& 0.993& 0.994& 0.992\\
  5& DLT & ExpPois & PoisConProp & 1.066& 1.093& 1.120& 1.148& 1.174& 1.120\\
  6& LT(22) & ExpPois & PoisConProp & 1.156& 1.215& 1.273& 1.338& 1.398& 1.276\\
 \hline
 \end{tabular}
 \caption{
Predictions of mean landfalling intense hurricane numbers.
 }\label{b16}
 \end{table}

 \begin{table}[h!]
   \centering
 \begin{tabular}{|c|c|c|c|c|c|c|c|c|c|}
  \hline
  1 & 2 & 3 & 4 & 5 & 6 & 7 & 8 & 9 & 10 \\
  \hline
 No. & SST Model & S2B Model & B2L Model &              2006 Var &                                                        2007 Var & 2008 Var & 2009 Var & 2010 Var                       & Mean Var \\
  \hline
  1& FL(8) & LinPois & PoisConProp & 1.216& 1.219& 1.220& 1.221& 1.221& 1.219\\
  2& DLT & LinPois & PoisConProp & 1.280& 1.301& 1.320& 1.338& 1.356& 1.319\\
  3& LT(22) & LinPois & PoisConProp & 1.344& 1.383& 1.419& 1.456& 1.491& 1.419\\
  4& FL(8) & ExpPois & PoisConProp & 1.353& 1.366& 1.368& 1.372& 1.373& 1.367\\
  5& DLT & ExpPois & PoisConProp & 1.465& 1.518& 1.564& 1.620& 1.664& 1.566\\
  6& LT(22) & ExpPois & PoisConProp & 1.616& 1.729& 1.838& 1.970& 2.080& 1.846\\
 \hline
 \end{tabular}
 \caption{
Prediction of the variance of landfalling intense hurricane numbers.
 }\label{b17}
 \end{table}

 \begin{table}[h!]
   \centering
 \begin{tabular}{|c|c|c|c|c|c|c|c|c|}
  \hline
  1 & 2 & 3 & 4 & 5 & 6 & 7 & 8 & 9 \\
  \hline
 No. & SST Model & SST2HU Model                       & 2006(T1,T2)                                                     & 2007(T1,T2)                                                     & 2008(T1,T2)                                                     & 2009(T1,T2)                                                     & 2010(T1,T2)                                                     \\
  \hline
  1& FL(8)  & LinPois & PoisConProp & 23, 77&23, 77&23, 77&23, 77&23, 77\\
  2& DLT    & LinPois & PoisConProp & 23, 77&23, 77&23, 77&23, 77&23, 77\\
  3& LT(22) & LinPois & PoisConProp & 23, 77&23, 77&23, 77&23, 77&23, 77\\
  4& FL(8)  & ExpPois & PoisConProp & 27, 73&27, 73&27, 73&28, 72&28, 72\\
  5& DLT    & ExpPois & PoisConProp & 27, 73&28, 72&28, 72&29, 71&29, 71\\
  6& LT(22) & ExpPois & PoisConProp & 28, 72&30, 70&31, 69&32, 68&33, 67\\
 \hline
 \end{tabular}
 \caption{
Breakdown of the variance in table~\ref{b17} into variance driven by SST prediction uncertainty and uncertainty in the SST-to-hurricane numbers regression model.
 }\label{b18}
 \end{table}

 \begin{table}[h!]
   \centering
 \begin{tabular}{|c|c|c|c|c|c|c|c|c|c|}
  \hline
  1 & 2 & 3 & 4 & 5 & 6 & 7 & 8 & 9 & 10 \\
  \hline
 No. & SST Model & S2B Model & B2L Model               & 2006 V/M &  2007 V/M & 2008 V/M & 2009 V/M & 2010 V/M          & Mean V/M \\
  \hline
  1& FL(8) & LinPois & PoisConProp & 1.298& 1.302& 1.302& 1.303& 1.303& 1.302\\
  2& DLT & LinPois & PoisConProp & 1.295& 1.298& 1.297& 1.297& 1.297& 1.297\\
  3& LT(22) & LinPois & PoisConProp & 1.293& 1.294& 1.293& 1.293& 1.291& 1.293\\
  4& FL(8) & ExpPois & PoisConProp & 1.369& 1.377& 1.379& 1.381& 1.381& 1.378\\
  5& DLT & ExpPois & PoisConProp & 1.375& 1.388& 1.397& 1.411& 1.417& 1.398\\
  6& LT(22) & ExpPois & PoisConProp & 1.397& 1.423& 1.443& 1.473& 1.487& 1.447\\
 \hline
 \end{tabular}
 \caption{
Ratio of the variance to the mean.
 }\label{b19}
 \end{table}

 \begin{table}[h!]
   \centering
 \begin{tabular}{|c|c|c|c|c|c|c|c|c|c|}
  \hline
  1 & 2 & 3 & 4 & 5 & 6 & 7 & 8 & 9 & 10 \\
  \hline
 No. & SST Model & S2B & B2L &                          2006 RMSE & 2007 RMSE & 2008 RMSE & 2009 RMSE & 2010 RMSE &       Mean\\
  \hline
  1& FL(8) & LinPois & PoisConProp & 0.241& 0.241& 0.242& 0.242& 0.242& 0.241\\
  2& DLT & LinPois & PoisConProp & 0.269& 0.278& 0.288& 0.298& 0.309& 0.289\\
  3& LT(22) & LinPois & PoisConProp & 0.301& 0.320& 0.340& 0.363& 0.388& 0.342\\
  4& FL(8) & ExpPois & PoisConProp & 0.299& 0.302& 0.302& 0.303& 0.303& 0.302\\
  5& DLT & ExpPois & PoisConProp & 0.352& 0.373& 0.393& 0.417& 0.442& 0.396\\
  6& LT(22) & ExpPois & PoisConProp & 0.421& 0.469& 0.522& 0.585& 0.654& 0.530\\
 \hline
 \end{tabular}
 \caption{
Standard errors on the predicted means.
 }\label{b20}
 \end{table}

\newpage
\begin{figure}[!hb]
  \begin{center}
    \scalebox{0.7}{\includegraphics{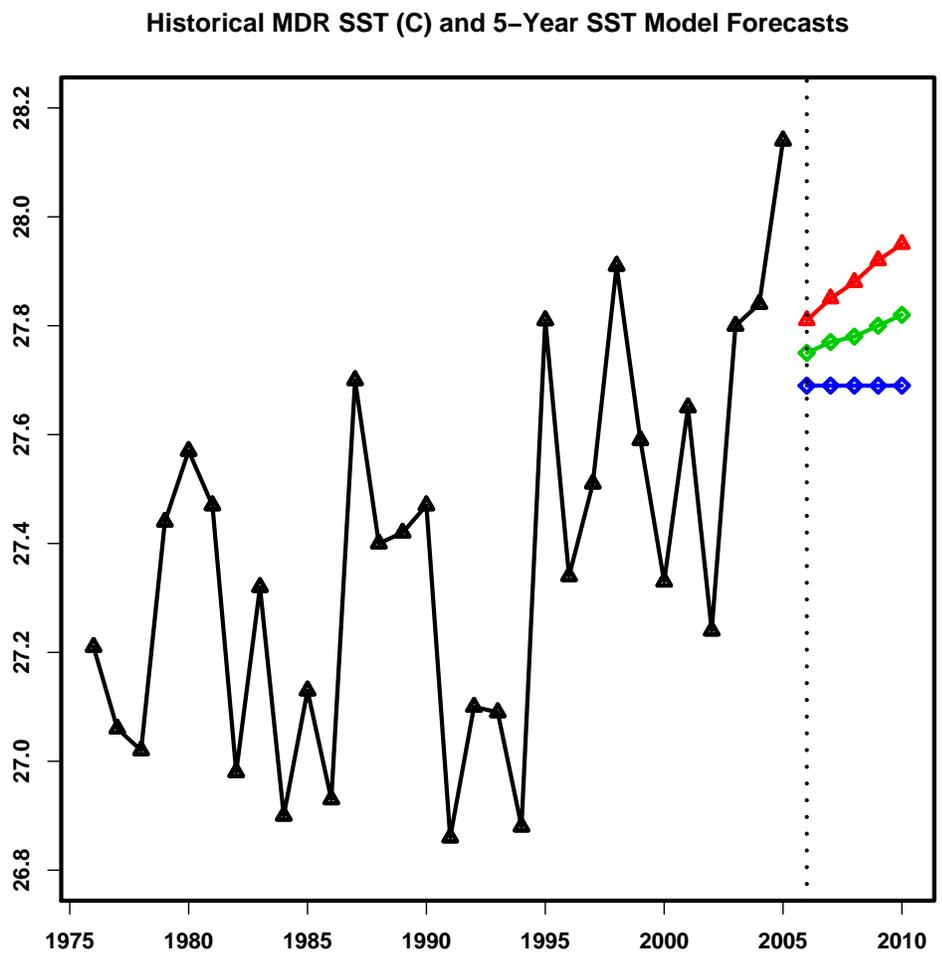}}
  \end{center}
    \caption{The three SST predictions we use as input to our hurricane prediction method, along
    with observed SSTs for the period 1976 to 2005.}
     \label{f01}
\end{figure}

\newpage
\begin{figure}[!hb]
  \begin{center}
    \scalebox{0.7}{\includegraphics{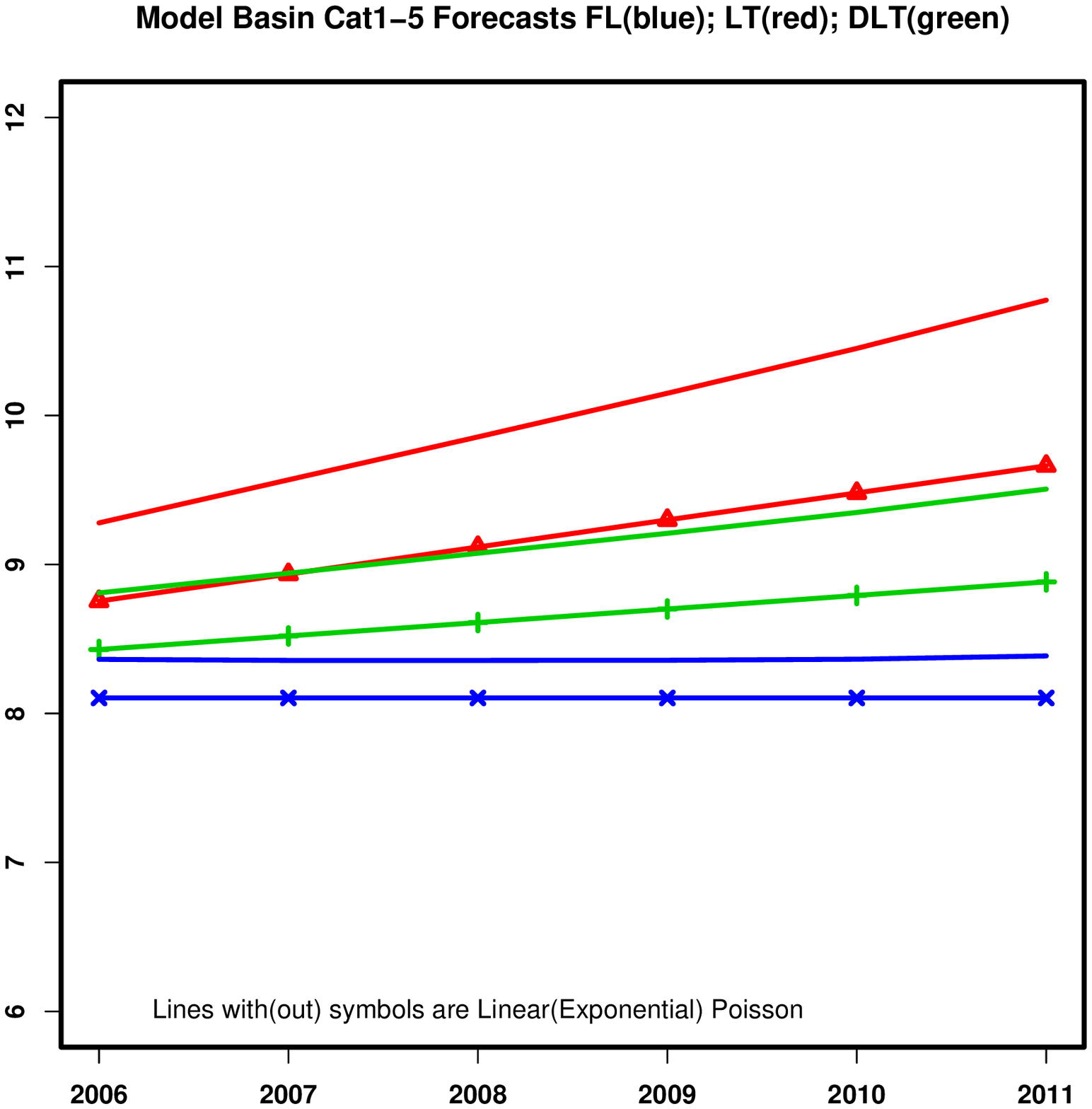}}
  \end{center}
    \caption{
Forecasts for the number of basin hurricanes for the years 2006 to
2011 for the 6 models described in the text. }
     \label{f02}
\end{figure}

\newpage
\begin{figure}[!hb]
  \begin{center}
    \scalebox{0.7}{\includegraphics{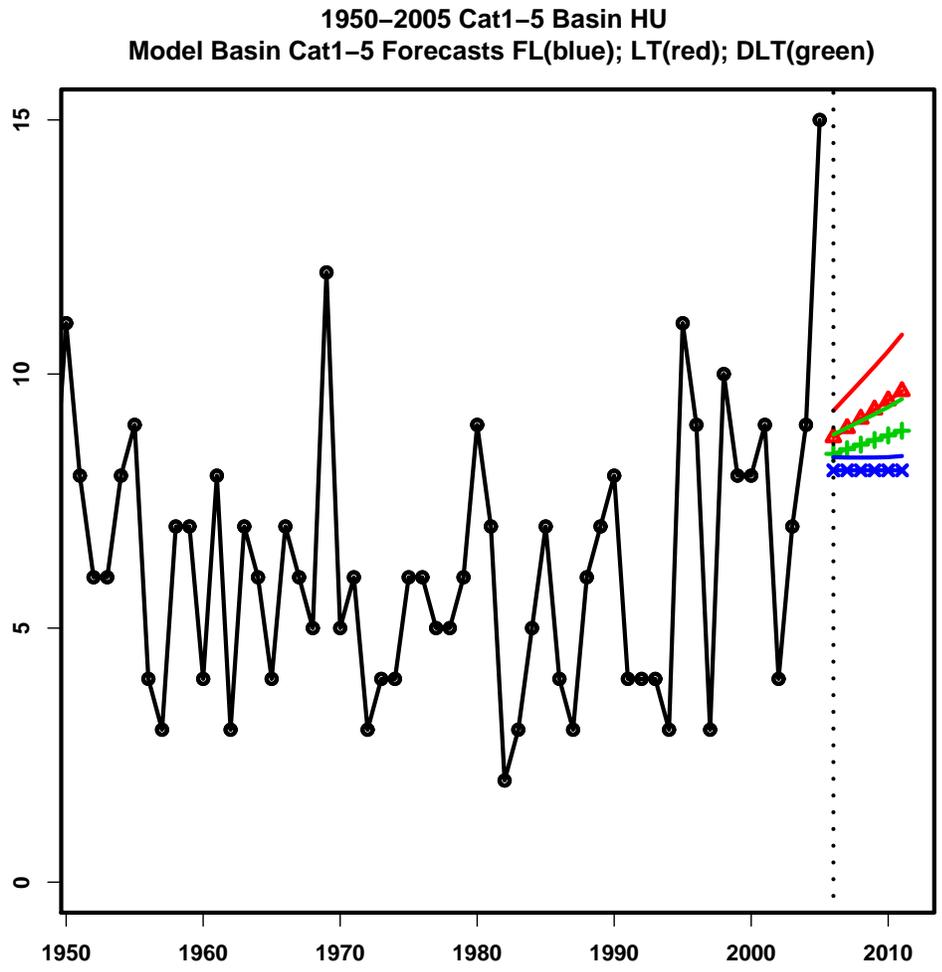}}
  \end{center}
    \caption{
As in figure~\ref{f02}, but also showing observed basin hurricane
numbers for the period 1950 to 2005. }
     \label{f03}
\end{figure}

\newpage
\begin{figure}[!hb]
  \begin{center}
    \scalebox{0.7}{\includegraphics{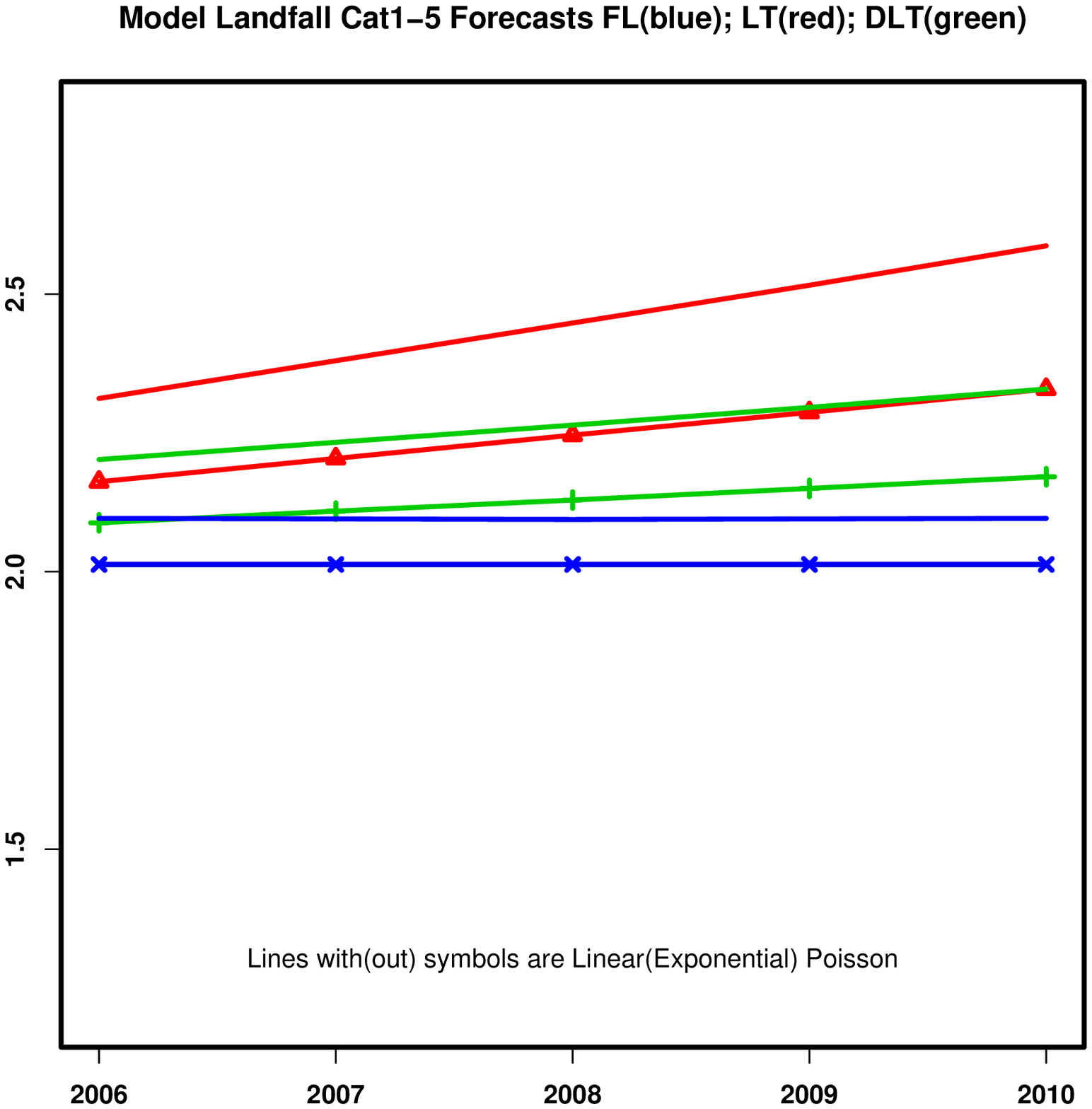}}
  \end{center}
    \caption{
Forecasts for the number of \emph{landfalling} hurricanes for the
years 2006 to 2011 for the 6 models described in the text. }
     \label{f04}
\end{figure}

\newpage
\begin{figure}[!hb]
  \begin{center}
    \scalebox{0.7}{\includegraphics{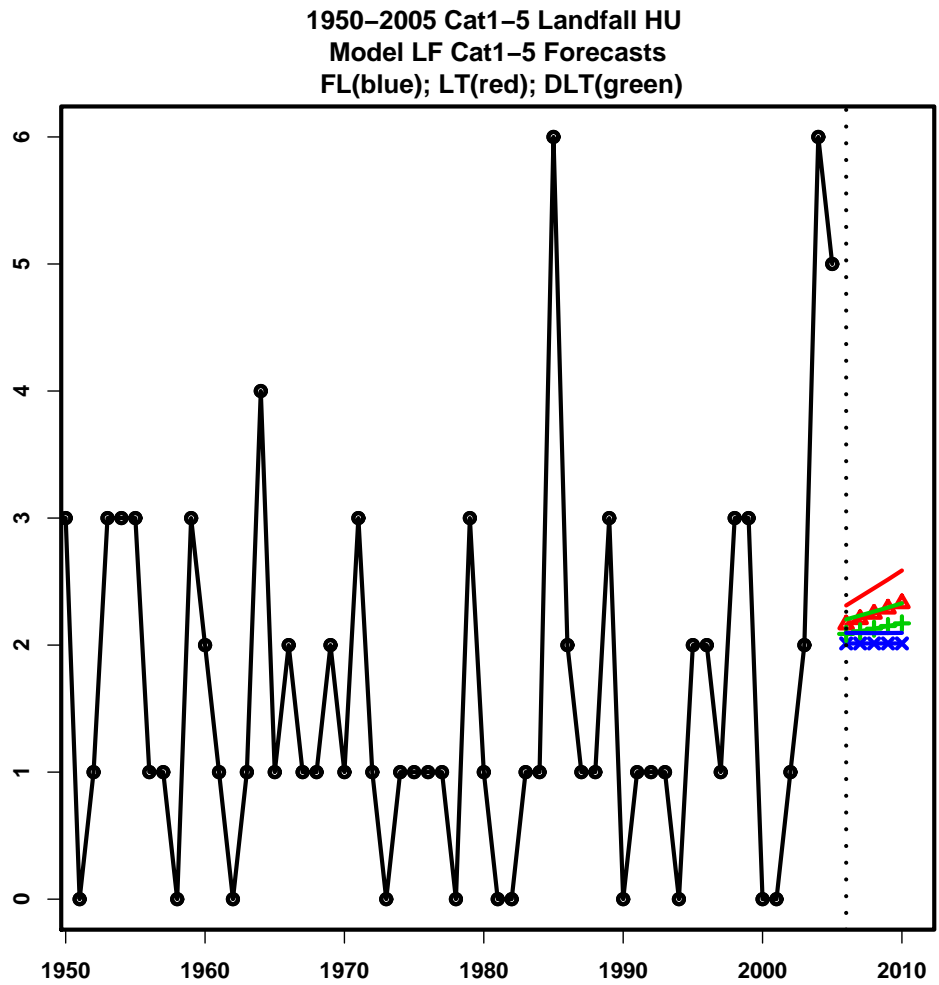}}
  \end{center}
    \caption{
As figure~\ref{f04}, but also showing observed landfalling
hurricane numbers for the period 1950 to 2005.}
     \label{f05}
\end{figure}

\newpage
\begin{figure}[!hb]
  \begin{center}
    \scalebox{0.7}{\includegraphics{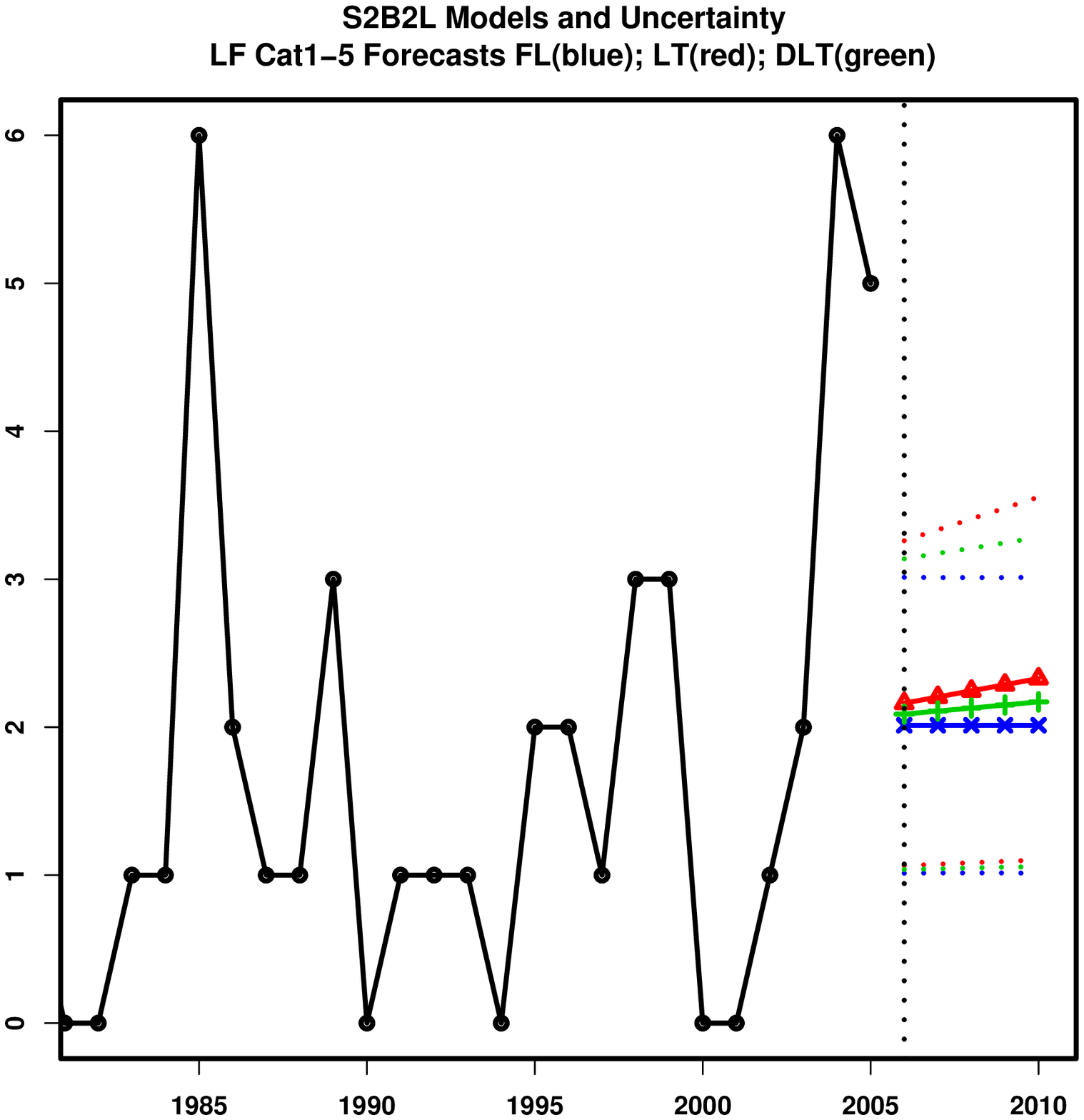}}
  \end{center}
    \caption{
As figure~\ref{f05}, but only showing observed data since 1980,
and the forecasts based on a linear conversion from SST to basin
hurricane numbers.}
     \label{f06}
\end{figure}

\newpage
\begin{figure}[!hb]
  \begin{center}
    \scalebox{0.7}{\includegraphics{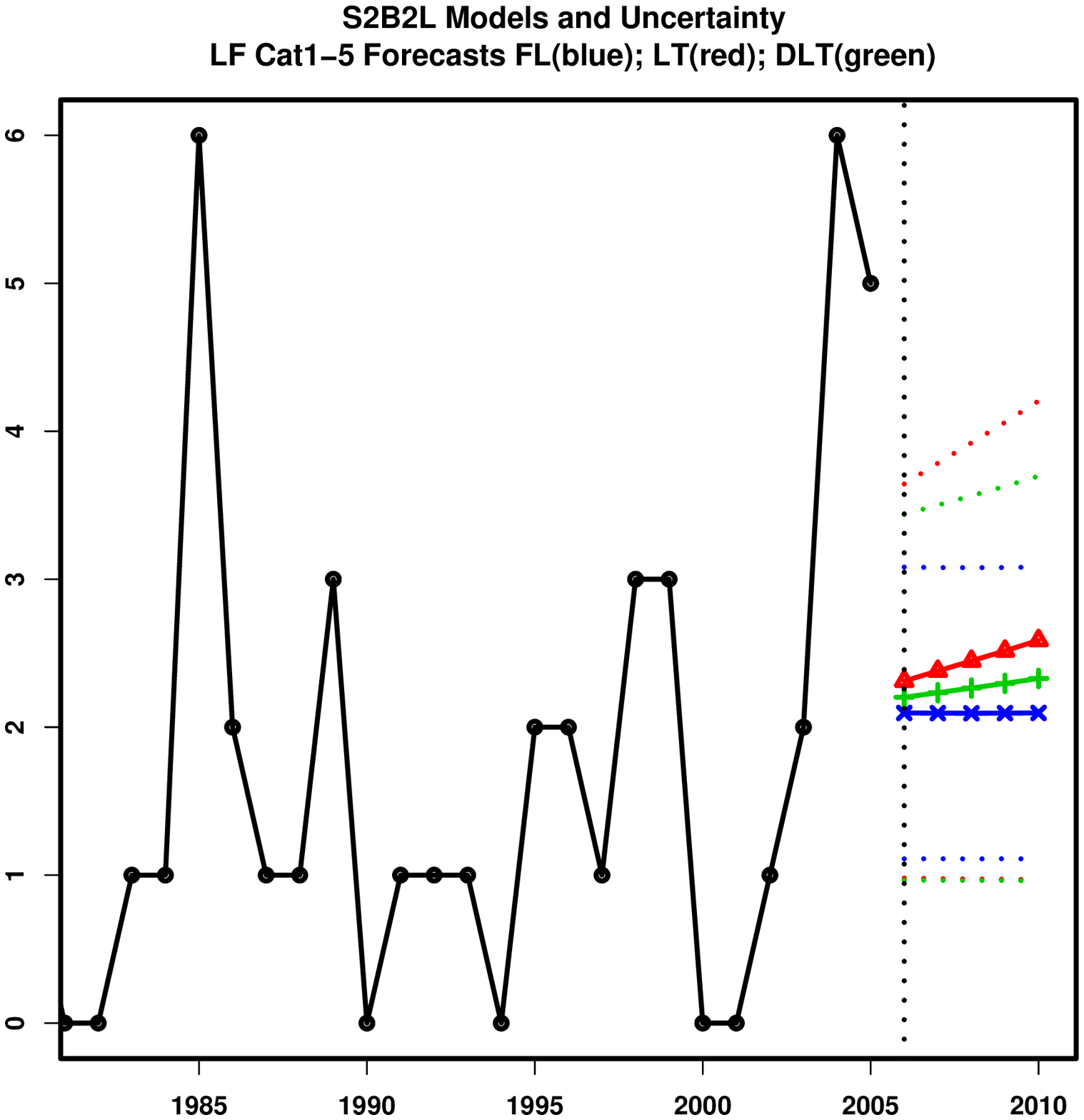}}
  \end{center}
    \caption{
As figure~\ref{f05}, but only showing observed data since 1980,
and the forecasts based on an \emph{exponential} conversion from
SST to basin hurricane numbers. }
     \label{f07}
\end{figure}

\end{document}